\documentclass[12pt]{iopart}
\usepackage{iopams}
\usepackage{setstack}
\newcommand{\half}{\mbox{$\textstyle \frac{1}{2}$}}
\newcommand{\quat}{\mbox{$\textstyle \frac{1}{4}$}}
\newcommand{\octa}{\mbox{$\textstyle \frac{1}{8}$}}
\newcommand{\toct}{\mbox{$\textstyle \frac{3}{8}$}}
\newcommand{\re}{{\rm e}}
\newcommand{\ri}{{\rm i}}
\newcommand{\rd}{{\rm d}}

\begin{document}
\title[Quantum noise and stochastic reduction]{Quantum noise and
stochastic reduction}

\author[Brody and Hughston]{Dorje~C~Brody$^{*}$ and
Lane~P~Hughston$^{\dagger}$}

\address{${}^{*}$Blackett Laboratory, Imperial College,
London SW7 2BZ, UK}

\address{${}^{\dagger}$Department of Mathematics, King's College
London, The Strand, London \\ WC2R 2LS, UK}

\begin{abstract}
In standard nonrelativistic quantum mechanics the expectation of
the energy is a conserved quantity. It is possible to extend the
dynamical law associated with the evolution of a quantum state
consistently to include a nonlinear stochastic component, while
respecting the conservation law. According to the dynamics thus
obtained, referred to as the energy-based stochastic Schr\"odinger
equation, an arbitrary initial state collapses spontaneously to
one of the energy eigenstates, thus describing the phenomenon of
quantum state reduction. In this article, two such models are
investigated: one that achieves state reduction in infinite time,
and the other in finite time. The properties of the associated
energy expectation process and the energy variance process are
worked out in detail. By use of a novel application of a nonlinear
filtering method, closed-form solutions---algebraic in character
and involving no integration---are obtained for both these models.
In each case, the solution is expressed in terms of a random
variable representing the terminal energy of the system, and an
independent noise process. With these solutions at hand it is
possible to simulate explicitly the dynamics of the quantum states
of complicated physical systems.
\end{abstract}

\submitto{\JPA}

%
%

\section{Introduction}
\label{sec:1}

The idea that the standard Schr\"odinger equation of
nonrelativistic quantum mechanics should be extended to take the
form of a stochastic differential equation on Hilbert space has
been investigated extensively as a mathematically viable approach
to the measurement problem in quantum mechanics. Indeed, there is
now a substantial body of literature on the theory of spontaneous
state vector reduction, and a number of different models have been
proposed that fall into this category. See, e.g., \cite{adler4,
ghirardi,pearle,percival4} for overviews of this area and relevant
references.

This paper is concerned with the so-called energy-based stochastic
extension of the Schr\"odinger equation, which has the special
status of being the simplest such extension that is in principle
applicable to any nonrelativistic quantum system. The physical
setup can be described briefly as follows. We consider an isolated
quantum system for which the Hamiltonian ${\hat H}$ has a discrete
spectrum $\{E_i\}_{i=1,2,\ldots,N}$. We assume that initially the
system is in a pure state represented by the state vector
$|\psi_0\rangle$. The situation where the initial state is mixed
can also be considered (see, e.g., \cite{brody1}), but for
simplicity we confine the discussion to the case of an initially
pure state in this paper. For convenience we set $\langle\psi_0
|\psi_0\rangle=1$. For each value of $i$ we let $|\phi_i\rangle$
denote the L\"uders state associated with the energy level $E_i$.
More specifically, let us write ${\hat\Pi}_i$ for the projection
operator onto the Hilbert subspace of states for which the energy
has the value $E_i$. We allow for the possibility that the energy
may be degenerate. In that case, we write $n_i$ for the number of
linearly independent state vectors with energy $E_i$. Then the
Hamiltonian takes the form ${\hat H}=\sum_i E_i{\hat\Pi}_i$, and
we define the L\"uders states by setting
\begin{eqnarray}
|\phi_i\rangle = \frac{{\hat\Pi}_i|\psi_0\rangle}{\langle\psi_0|
{\hat\Pi}_i|\psi_0\rangle^{1/2}}.
\end{eqnarray}
We note that ${\hat H}|\phi_i\rangle=E_i|\phi_i\rangle$ and
$\langle \phi_i |\phi_i\rangle=1$. According to the von
Neumann-L\"uders projection postulate~\cite{luders,vonneuman}, if
the system is initially in the pure state $|\psi_0\rangle$ and if
the outcome of a measurement of the energy is the eigenvalue
$E_i$, then after the measurement the state of the system will be
the L\"uders state $|\phi_i\rangle$.

It is an implicit feature of the projection postulate that quantum
evolution progresses in accordance with the unitary dynamics of
the Schr\"odinger equation up to the moment when a measurement is
made, at which point the system jumps to a new state. The von
Neumann-L\"uders rule asserting how the jump proceeds is
essentially {\em ad hoc} in nature, despite being plausible from a
physical point of view insofar as the predicted outcome is
concerned. Thus although the projection postulate, in one form or
another, remains an accepted part of the everyday use of quantum
theory in practical applications~\cite{isham}, one has to agree
that such a `cookbook' approach to the measurement problem is
ultimately unsatisfactory; and this is why over the last five
decades many attempts have been made to modify the dynamics of
standard quantum mechanics in such a way that the `collapse'
process can be understood as governed by an evolutionary law that
operates on a universal basis, rather as does the Schr\"odinger
equation in ordinary quantum mechanics.

In order to ensure consistency with established facts, such a
universal evolutionary law needs to have the property that for
some systems it proceeds in a way that for all practical purposes
reproduces the dynamics of the Schr\"odinger equation, whereas for
other systems the evolution progresses continuously to a terminal
state that is consistent with the action of the projection
postulate. This `viability' property is satisfied in particular by
the standard energy-based stochastic Schr\"odinger equation. In
this model the Schr\"odinger equation, which when written in
differential form is given by
\begin{eqnarray}
{\rm d}|\psi_t\rangle = -{\rm i}{\hat H} |\psi_t\rangle {\rm d}t,
\end{eqnarray}
is generalised and elevated to the status of a nonlinear
stochastic differential equation on Hilbert space:
\begin{eqnarray}
{\rm d}|\psi_t\rangle = -{\rm i}{\hat H} |\psi_t\rangle {\rm d}t -
\octa \sigma^2({\hat H}-H_t)^2 |\psi_t\rangle {\rm d}t + \half
\sigma ({\hat H}-H_t)|\psi_t\rangle {\rm d}W_t. \label{eq:1.1}
\end{eqnarray}
Here $\{|\psi_t\rangle \}_{0\leq t<\infty}$ is the state-vector
process, $\{W_t\}_{0\leq t<\infty}$ is the Wiener process, and
$\{H_t\}_{0\leq t<\infty}$ is the energy expectation process,
defined by
\begin{eqnarray}
H_t = \frac{\langle{\psi}_t|{\hat H}|\psi_t\rangle}
{\langle{\psi}_t|\psi_t\rangle}. \label{eq:1.2}
\end{eqnarray}
The coupling constant $\sigma$ appearing in (\ref{eq:1.1}), which
has the units $ \sigma\sim[{\rm energy}]^{-1}[{\rm time}]^{-1/2}$,
determines the characteristic timescale $\tau_R$ associated with
the rate of collapse of the wave function induced by
(\ref{eq:1.1}). This timescale is given typically by an expression
of the form $\tau_R = 1/\sigma^2 V_0$, where $V_0$ is the initial
value of the process $\{V_t\}_{0\leq t<\infty}$ for the energy
variance, which at time $t$ is given by the following expression:
\begin{eqnarray}
V_t = \frac{\langle{\psi}_t|({\hat H}-H_t)^2|\psi_t\rangle}
{\langle{\psi}_t|\psi_t\rangle}. \label{eq:1.5}
\end{eqnarray}

One of the attractive features of the stochastic differential
equation (\ref{eq:1.1}) is that it provides a more or less
completely tractable model for state vector reduction in
nonrelativistic quantum mechanics. Nevertheless, despite the fact
that the mathematical properties and physical consequences of
(\ref{eq:1.1}) and various related processes have been studied
extensively in the literature~\cite{adler3,adler,brody1,adler,
brody2,diosi,gisin,grw,hughston,percival}, it has only been
recently that a general solution of (\ref{eq:1.1}) has been
obtained in terms of an appropriate set of freely specifiable
random data ~\cite{brody3,brody4}. The aim of this paper is to
present a complete treatment of the method of solution of the
dynamical equation (\ref{eq:1.1}). The results are of interest
both for the new range of numerical and computational techniques
they open up, as well as for the new methods for model building
they provide.

The article is organised as follows. In \S\ref{sec:2} we review
some basic notions of stochastic analysis, including the concepts
of filtrations, conditional expectations, martingales,
supermartingales, and potentials. The material introduced in this
section will be used throughout the paper. In \S\ref{sec:3},
\S\ref{sec:4}, and \S\ref{sec:5} we establish various properties
of the energy expectation process (\ref{eq:1.2}) and the energy
variance process (\ref{eq:1.5}), showing that the variance process
has the `potential' property; that is to say, its expectation goes
to zero asymptotically. This allows us to give a precise sense to
the notion of state reduction. In \S\ref{sec:6} we then determine
the circumstances under which the state vector reduces to one of
the eigenstates of an observable that is compatible with the
Hamiltonian.

In \S\ref{sec:7} we address the problem of the origin of the
dynamical equation (\ref{eq:1.1}). Starting from a general
stochastic equation for a state vector driven by a single Brownian
motion, we determine what additional physical assumptions and
other simplifying features are required in order to obtain
(\ref{eq:1.1}). We also show, under a suitable `universality'
assumption regarding energy conservation, that reduction to lower
energy uncertainty is a generic feature of the stochastic
Schr\"odinger equation.

The projection operators for the energy eigenstates constitute a
special set of observables that commute with the Hamiltonian. The
expectation value, with respect to the state $|\psi_t\rangle$, of
such a projection operator determines the random process for the
associated conditional transition probability to that eigenstate.
The properties of this conditional probability process are studied
in \S\ref{sec:8}. In \S\ref{sec:9} we digress briefly to
investigate the dynamics of the Shannon entropy associated with
the system of transition probabilities, and show that the Shannon
entropy has the property that its expectation goes to zero
asymptotically. This is contrasted with the behaviour of the von
Neumann entropy. We also derive an inequality that relates the
entropic measure and the variance-based measure of energy
dispersion. In \S\ref{sec:10} we study a certain linear stochastic
differential equation for the state vector, which we call the
ancillary equation, and verify that a state vector satisfying the
ancillary equation, once suitably normalised, can be used as a
step to obtaining the solution of the nonlinear equation
(\ref{eq:1.1}). We also clarify the relation of our results to
earlier work on solutions of (\ref{eq:1.1}) and related dynamical
equations, explaining why previously established integral
representations for the state vector satisfying the stochastic
equation should not generally be regarded as explicit solutions in
the sense that we use the term here.

Then in \S\ref{sec:11} and \S\ref{sec:12} we derive a bona fide
explicit solution to (\ref{eq:1.1}), making use of a nonlinear
filtering method. The solution thus obtained is expressed in terms
of a simple algebraic function of a standard Brownian motion and
an independent random variable representing the terminal value of
the energy. By use of this result, it is possible to simulate
solutions that represent the evolution of rather complicated
quantum systems. In \S\ref{sec:13} we introduce a technique that
allows one to verify that the solution obtained in \S\ref{sec:11}
does indeed give rise to the reduction of the state vector. In
\S\ref{sec:14} we investigate properties of the asymptotic random
variable corresponding to the terminal value of the energy. While
in the nonlinear filtering method used to solve (\ref{eq:1.1}) we
introduce a noise term, in \S\ref{sec:15} we derive the external
noise term from the underlying processes specified in
(\ref{eq:1.1}). This result justifies the use of the filtering
methodology we have employed here.

We then turn to solve the problem of constructing a collapse model
that achieves state reduction in finite time. That is, although
the standard energy-based collapse model (\ref{eq:1.1}) achieves a
strict collapse in infinite time, with a minimal modification of
the dynamical equation (\ref{eq:1.1}) it is possible to formulate
a finite-time collapse model. In \S\ref{sec:16} we introduce such
a model. Using the methodology of \S\ref{sec:11} we derive an
analytical expression that we conjecture to give the energy
expectation process. The validity of this conjecture is
established in \S\ref{sec:17}. In \S\ref{sec:18} we derive the
external noise term arising in the finite-time collapse model that
is used in \S\ref{sec:17} to obtain the solution. In
\S\ref{sec:19} we demonstrate the fact that the standard
energy-based model (\ref{eq:1.1}) and the finite-time collapse
model introduced in (\ref{eq:16.1}) are related by a nonlinear
time-change. That is, if we take the model (\ref{eq:1.1}) and
replace the time variable $t$ by a `clock' variable defined by
$\tau(t)=tT/(T-t)$, where $T$ is a finite positive constant, then
in a physical world measured by the variable $t$, the collapse for
the new system takes place in finite time interval $T$, since the
clock variable $\tau(t)$ runs from 0 to $\infty$ as $t$ runs from
$0$ to $T$.

As a closing remark, in \S\ref{sec:20}, the role of the asymptotic
value of the energy, which has the interpretation of a hidden
variable in the stochastic framework, is discussed. We also
speculate on whether the energy-based reduction models analysed
here suffice as such to form a basis for the general description
of random phenomena in nonrelativistic quantum mechanics.

\section{Stochastic essentials}
\label{sec:2}

We begin with an overview of the probabilistic framework implicit
in the specification of the energy-based stochastic Schr\"odinger
equation. The concepts introduced in this section are standard in
the literature of stochastic analysis, as is also the notation
(see, e.g., \cite{karatzas,shiryaev,protter,revuz,williams,yor}).
The dynamics of the state vector $|\psi_t\rangle$ are defined on a
probability space $(\Omega,{\mathcal F},{\mathbb P})$ with
filtration $\{{\mathcal F}_t\}_{0\leq t<\infty}$, with respect to
which the process $\{W_t\}_{0\leq t<\infty}$ is a standard
Brownian motion. Here $\Omega$ is the sample space on which
${\mathcal F}$ is a $\sigma$-algebra of open sets upon which the
probability measure ${\mathbb P}$ is defined. Each element
$\omega\in\Omega$ represents a `possible outcome of chance'. Each
element $A\in{\mathcal F}$ is an `event'. The measure ${\mathbb
P}$ assigns a probability ${\mathbb P}(A)$ to each event $A$.

Now we give the relevant definitions in more detail, since these
are of interest. Let $\Omega$ be a set, and let ${\mathcal F}$ be
a collection of subsets of $\Omega$. For any subset $A\subset
\Omega$ we let $A^{\rm c}=\{\omega\in\Omega|\omega\notin A\}$
denote its complement. Then ${\mathcal F}$ is called an algebra of
subsets of $\Omega$ if (a) $\Omega\in{\mathcal F}$, (b)
$A\in{\mathcal F}$ implies that $A^{\rm c} \in {\mathcal F}$, and
(c) $A,B\in{\mathcal F}$ implies $A\cup B \in {\mathcal F}$. It
follows from these axioms that $\emptyset\in{\mathcal F}$, and
that $A,B\in{\mathcal F}$ implies $A\cap B\in{\mathcal F}$.

The algebraic operations on the elements of ${\mathcal F}$ are as
follows. The product of two elements $A,B\in{\mathcal F}$ is
defined by $A\cdot B=A\cap B$, and the sum of two elements
$A,B\in{\mathcal F}$ is defined by $A+ B=(A\cup B)\cap(A\cap
B)^{\rm c}$. It follows that the product and sum operations are
symmetric and associative, and that $(A+B) \cdot C=A\cdot C+B\cdot
C$ for any $A,B,C\in{\mathcal F}$. The underlying field of the
algebra ${\mathcal F}$ is the minimal subalgebra
$\{\Omega,\emptyset\}$, which when endowed with the same product
and sum operations as those defined above satisfies the rules of
binary arithmetic:
\begin{eqnarray}
\left\{ \begin{array}{lll} \emptyset\cdot\emptyset=\emptyset, &
\emptyset\cdot\Omega=\emptyset, & \Omega\cdot\Omega=\Omega, \\
\emptyset+\emptyset=\emptyset, & \emptyset+\Omega=\Omega, &
\Omega+\Omega=\emptyset. \end{array} \right.
\end{eqnarray}

If ${\mathcal F}$ is an algebra of subsets of $\Omega$ then we say
that ${\mathcal F}$ is a $\sigma$-algebra if it has the property
that whenever $\{A_n\}_{n\in{\mathbb N}}\in{\mathcal F}$ then
$\cup_n A_n\in{\mathcal F}$. That is to say, the union of any
countable sequence of elements of ${\mathcal F}$ is also an
element of ${\mathcal F}$. It follows that whenever $\{A_n \}_{
n\in{\mathbb N}}\in {\mathcal F}$ we have $\cap_n A_n \in{\mathcal
F}$, since $\cap_n A_n=(\cup_n A_n^{\rm c})^{\rm c}$.

We comment briefly on the distinction between a $\sigma$-algebra
and a topology, since the latter is more familiar to physicists
than the former. In a topology we axiomatise the notion of an open
set, and require that the union of any collection of open sets is
open, and that the intersection of any finite collection of open
sets is open. In a $\sigma$-algebra we axiomatise the notion of a
measurable set, and require that the union of any countable
sequence of measurable sets is measurable, and that the
intersection of any countable sequence of measurable sets is
measurable.

If ${\mathcal F}$ is a $\sigma$-algebra of subsets of a set
$\Omega$, then we call the pair $(\Omega,{\mathcal F})$ a
measurable space. If $(\Omega,{\mathcal F})$ is a measurable
space, then a probability measure on $(\Omega,{\mathcal F})$ is a
map ${\mathbb P}:\ {\mathcal F}\to[0,1]$ satisfying: (a) ${\mathbb
P}(\emptyset)=0$, (b) ${\mathbb P}(\Omega)=1$, and (c) if
$\{A_n\}_{n\in{\mathbb N}}$ is a countable sequence of disjoint
elements of ${\mathcal F}$ with union $A= \cup_nA_n$, then
${\mathbb P}(A)=\sum_n{\mathbb P}(A_n)$. A triple
$(\Omega,{\mathcal F},{\mathbb P})$ is called a probability space.

The introduction of the concept of a filtration on a probability
space allows one to formalise the notion that the consequences of
the outcome of chance are not necessarily revealed at once, but
rather may emerge sequentially as time progresses. More
specifically, a filtration of ${\mathcal F}$ is a collection
$\{{\mathcal F}_t\}$ of $\sigma$-subalgebras of ${\mathcal F}$
such that ${\mathcal F}_s\subset {\mathcal F}_t$ for all $s$ and
$t$ such that $0\leq s\leq t<\infty$.

If an event $A\in{\mathcal F}$ is such that $A\in{\mathcal F}_t$
for some given value of $t$, then we interpret this to mean that
at time $t$ one can say whether $\omega\in A$ or not. To put this
another way, in a filtered probability space each $\omega \in
\Omega$ corresponds to a possible `future history'. Each element
$A\in{\mathcal F}_t$ then represents a simple yes/no question, the
answer to which, for any particular future history, will be known
for certain by time $t$. For that reason, the nesting ${\mathcal
F}_s\subset{\mathcal F}_t$ for $s\leq t$ gives rise to a notion of
causality.

A real-valued function $X:\ \Omega\mapsto{\mathbb R}$ is said to
be \emph{measurable} with respect to the $\sigma$-algebra
${\mathcal F}$ if for each number $x\in{\mathbb R}$ the set
$\{\omega\in\Omega: {\bf 1}_x (\omega)=1\}$ is an element of
${\mathcal F}$. Here ${\bf 1}_x (\omega)$ is the indicator
function on $\Omega$ for the set consisting of all $\omega$ such
that $X(\omega)\leq x$. Thus ${\bf 1}_x: \Omega\mapsto\{0,1\}$,
and ${\bf 1}_x(\omega)={\bf 1}_{\{X(\omega)\leq x\}}$. If $X$ is
${\mathcal F}$-measurable in the sense just discussed, we say that
$X$ is a real-valued random variable on $(\Omega,{\mathcal
F},{\mathbb P})$. The probability distribution function
$F_X(x)={\mathbb P}(X\leq x)$ is then defined by use of the
Lebesgue integral:
\begin{eqnarray}
{\mathbb P}(X\leq x) = \int_{\Omega} {\bf 1}_x(\omega) {\rm
d}{\mathbb P}(\omega) .
\end{eqnarray}
More generally, we also consider maps of the form $X:\ \Omega
\mapsto {\Gamma}$ where $(\Omega,{\mathcal F},{\mathbb P})$ is a
probability space and $(\Gamma,{\mathcal G})$ is a measurable
space. For example, $\Gamma$ could be ${\mathbb R}^n$, ${\mathbb
C}^n$, a Hilbert space, or a manifold. In that case we say the
random variable $X$ takes values in $\Gamma$, and ${\mathcal G}$
can be typically taken to be the so-called Borel $\sigma$-algebra
generated by the open sets of $\Gamma$. Then for any element $G\in
{\mathcal G}$ we define
\begin{eqnarray}
{\mathbb P}(X\in G) = \int_\Omega {\bf 1}_{\{X(\omega)\in G\}} \rd
{\mathbb P}(\omega).
\end{eqnarray}

A parametric family $\{X_t\}_{0\leq t<\infty}$ of random variables
on $(\Omega,{\mathcal F},{\mathbb P})$ is called a random process.
If a random process $\{X_t\}$ on a probability space
$(\Omega,{\mathcal F},{\mathbb P})$ with filtration $\{{\mathcal
F}_t\}$ has the property that for each value of $t$ the random
variable $X_t$ is ${\mathcal F}_t$-measurable, then we say that
$\{X_t\}$ is \emph{adapted} to the filtration $\{{\mathcal
F}_t\}$.

If $X$ is a nonnegative real random variable, its ${\mathbb
P}$-expectation (i.e. its expectation with respect to the measure
${\mathbb P}$) is defined by the integral
\begin{eqnarray}
{\mathbb E}[X] = \int_{\Omega} X(\omega) {\rm d}{\mathbb P}
(\omega) ,
\end{eqnarray}
which may take the value $+\infty$. More generally, when $X$ is
not necessarily nonnegative, the expectation is defined only when
one of the expressions ${\mathbb E}[X^+]$ or ${\mathbb E}[X^-]$ is
finite, where $X^+=\max(X,0)$ and $X^-=-\min(X,0)$, in which case
${\mathbb E}[X] = {\mathbb E} [X^+]-{\mathbb E}[X^-]$. A random
variable such that ${\mathbb E} [|X|]={\mathbb E}[X^+]+{\mathbb
E}[X^-]$ is finite is said to be {\it integrable}.

Now we turn to the definition of conditional expectation. Given a
random variable $X$ on $(\Omega,{\mathcal F},{\mathbb P})$ for
which ${\mathbb E}[X]$ exists, the conditional expectation
${\mathbb E}[X|{\mathcal A}]$ of $X$ with respect to the
$\sigma$-subalgebra ${\mathcal A}\subset{\mathcal F}$ is defined
to be any ${\mathcal A}$-measurable random variable $Y$ for which
${\mathbb E}[Y]$ is defined, such that for any element
$A\in{\mathcal A}$ we have
\begin{eqnarray}
\int_{\Omega} {\bf 1}_A(\omega) X(\omega) {\rm d}{\mathbb P}
(\omega) = \int_{\Omega} {\bf 1}_A(\omega) Y(\omega) {\rm d}
{\mathbb P} (\omega) .
\end{eqnarray}
If such a random variable exists, then it is unique up to
equivalence modulo differences on sets of ${\mathbb P}$-measure
zero. Thus even if ${\mathbb E}[X|{\mathcal A}]$ is not quite
unique we refer to it as {\it the} conditional expectation of $X$
with respect to ${\mathcal A}$. This definition, which at first
glance appears rather formal and indirect, is nevertheless one of
the cornerstones of modern probability theory, and is
indispensable. We remark that a sufficient condition for ${\mathbb
E}[X|{\mathcal A}]$ to exist is that $X$ should be integrable.

The following properties of the conditional expectation are often
useful in calculations: (i) the law of total probability ${\mathbb
E}[{\mathbb E}[X|{\mathcal A}]] = {\mathbb E}[X]$; and (ii) the
tower property, which says that if ${\mathcal A} \subset{\mathcal
B}\subset{\mathcal F}$ then ${\mathbb E}[ {\mathbb E}[X|{\mathcal
B}]|{\mathcal A}]={\mathbb E}[X|{\mathcal A}]$. The law of total
probability is a special case of the tower property.

The conditional expectation operation allows us to introduce the
concept of a \emph{martingale}, the stochastic analogue of a
conserved quantity. For this purpose we need the operation of
conditioning with respect to a $\sigma$-subalgebra ${\mathcal
F}_t$ belonging to a filtration $\{{\mathcal F}_t\}_{0\leq
t<\infty}$. Intuitively, conditioning with respect to ${\mathcal
F}_t$ means conditioning with respect to the information that will
become available up to time $t$. For convenience, we often use the
abbreviation ${\mathbb E}_t[X]={\mathbb E}[X|{\mathcal F}_t]$ when
the choice of filtration can be taken as understood. There are
situations, however, where more than one filtration may arise in
the context of a given problem, in which case the more explicit
notation is useful. The conditional expectation ${\mathbb E}_t[X]$
satisfies ${\mathbb E}[{\mathbb E}_t[X]] ={\mathbb E}[X]$ and
${\mathbb E}_s[{\mathbb E}_t [X]]={\mathbb E}_s[X]$ for $s\leq t$.
We note that if $X$ is ${\mathcal F}_t$-measurable, then ${\mathbb
E}_t[X]=X$.

A real-valued process $\{X_t\}$ is said to be an $\{{\mathcal
F}_t\}$-{\em martingale} on the probability space
$(\Omega,{\mathcal F},{\mathbb P})$ if ${\mathbb E}[|X_t|]<\infty$
for all $0\leq t<\infty$, and ${\mathbb E}_s[X_t]=X_s$ for all
$0\leq s\leq t<\infty$. In other words, $\{X_t\}$ is an
$\{{\mathcal F}_t\}$-martingale if it is integrable and if for
$t\geq s$ the conditional expectation of $X_t$, given ${\mathcal
F}_s$, is the value $X_s$ of the process at time $s$. A process
$\{X_t\}$ is an $\{{\mathcal F}_t\}$-{\em supermartingale} on
$(\Omega,{\mathcal F},{\mathbb P})$ if ${\mathbb E}[|X_t|]<\infty$
for all $t\geq0$, and ${\mathbb E}_s[X_t]\leq X_s$ for all $0\leq
s\leq t<\infty$. Intuitively, a supermartingale is a process that
tends on average, at any time, to be nonincreasing. A martingale
is {\it a fortiori} a supermartingale. The martingale convergence
theorem (see, e.g., \cite{protter}, theorem~10) states that if
$\{X_t\}$ is a supermartingale that satisfies
\begin{eqnarray}
\sup_{0\leq t<\infty}{\mathbb E}[|X_t|]<\infty,
\end{eqnarray}
then there exists a random variable $Y$ such that
$\lim_{t\to\infty}X_t=Y$ almost surely (i.e. with probability
one), and that ${\mathbb E}[|Y|]<\infty$. It follows that a
positive supermartingale necessarily converges to a limit as $t$
goes to infinity. A positive supermartingale $\{X_t\}$ with the
property that $\lim_{t\to\infty}{\mathbb E}[X_t]=0$ is called a
{\sl potential}~\cite{meyer}.

We now review some basic formulae arising in the theory of Ito
processes. Let $(\Omega,{\mathcal F},{\mathbb P})$ be a
probability space with filtration $\{{\mathcal F}_t\}_{ 0\leq
t<\infty}$, and let $\{W_t\}_{ 0\leq t<\infty}$ be a standard
Wiener process adapted to $\{{\mathcal F}_t\}$. Here by a standard
Wiener process (or Brownian motion) we mean a continuous process
$\{W_t\}$ with the properties that (i) $\{W_t\}$ has independent
increments, and that (ii) $W_t-W_s$ for $0\leq s<t<\infty$ is a
Gaussian random variable with mean zero and variance $t-s$.

Let $\{a_t\}_{ 0\leq t<\infty}$ and $\{b_t\}_{ 0\leq t<\infty}$ be
$\{{\mathcal F}_t\}$-adapted processes such that for any
$t\in[0,\infty)$ we have
\begin{eqnarray}
\int_0^t |a_s|\rd s + \int_0^t b_s^2 \rd s < \infty
\end{eqnarray}
almost surely. Then letting $X_0$ an ${\mathcal F}_0$-measurable
initial condition, the random variable $X_t$ defined by the
stochastic integral
\begin{eqnarray}
X_t = X_0 + \int_0^t a_s\rd s + \int_0^t b_s \rd W_s
\label{eq:stc}
\end{eqnarray}
is well defined and ${\mathcal F}_t$-measurable for all $t$, and
we call $\{X_t\}_{ 0\leq t<\infty}$ an Ito process (see, e.g.,
\cite{protter,revuz} for the general definition of stochastic
integration). In this case we say that $\{X_t\}$ is a real-valued
Ito process driven by the one-dimensional Wiener process
$\{W_t\}$. It is straightforward to generalise (\ref{eq:stc}) to
cases for which both $\{X_t\}$ and $\{W_t\}$ are multidimensional.

One useful tool of which we make repeated use is Ito's lemma.
Suppose $\{X_t\}$ is given by (\ref{eq:stc}) and consider the
process $\{f_t\}_{0\leq t<\infty}$ defined by $f_t=f(X_t,t)$ where
$f\in C^{2,1}({\mathbb R}\times{\mathbb R}^+)$. Let prime and dot
denote differentiation with respect to the first and second
arguments of $f(x,t)$, respectively. Then Ito's lemma states that
\begin{eqnarray}
f(X_t,t)&=&f(X_0,0)+\int_0^t {\dot f}(X_s,s)\rd s+ \int_0^t a_s
f'(X_s,s)\rd s \nonumber \\ && + \half\int_0^t b_s^2
f^{\prime\prime}(X_s,s)\rd s + \int_0^t b_s f^\prime(X_s,s) \rd
W_s. \label{eq:stc2}
\end{eqnarray}
It is often convenient to express (\ref{eq:stc}) and
(\ref{eq:stc2}) in differential form: thus we write
\begin{eqnarray}
\rd X_t = a_t \rd t + b_t \rd W_t \label{eq:sde}
\end{eqnarray}
for the `dynamics' of $\{X_t\}$, and
\begin{eqnarray}
\fl \hspace{1.0cm} \rd f(X_t,t)= \left( {\dot f}(X_t,t) + a_t
f'(X_t,t)+\half b_t^2 f^{\prime\prime}(X_t,t)\right)\rd t + b_t
f^\prime(X_t,t) \rd W_t \label{eq:stc2-5}
\end{eqnarray}
for the dynamics of $\{f_t\}$ implied by Ito's lemma. As in
ordinary calculus, the differential equations of stochastic
calculus are essentially formal in character, and always derive
their meaning from associated integral equations. Thus
(\ref{eq:sde}) and (\ref{eq:stc2-5}) refer back to (\ref{eq:stc})
and (\ref{eq:stc2}). Nevertheless, as in ordinary calculus, the
manipulation of infinitesimal quantities in stochastic calculus
can be very powerful as a mathematical technique, and can be
intuitively very suggestive as well. For example, the so-called
Ito product rule
\begin{eqnarray}
\rd(X_tY_t)=Y_t\rd X_t+X_t\rd Y_t+\rd X_t \rd Y_t
\end{eqnarray}
is short-hand for the fact that if $\{X_t\}$ is given by
(\ref{eq:stc}) and $\{Y_t\}$ is given analogously, but with
$\{a_t\}$ and $\{b_t\}$ replaced by $\{p_t\}$ and $\{q_t\}$, then
\begin{eqnarray}
X_tY_t=X_0Y_0+\int_0^t(Y_sa_s+X_sp_s+b_sq_s)\rd s+\int_0^t(Y_sb_s+
X_sq_s)\rd W_s.
\end{eqnarray}

Now consider an Ito process $\{M_t\}_{0\leq t<\infty}$ of the form
\begin{eqnarray}
M_t = M_0 + \int_0^t b_s \rd W_s.
\end{eqnarray}
Then a sufficient condition for $\{M_t\}$ to be a martingale is
that $M_0$ should be integrable, and that
\begin{eqnarray}
{\mathbb E}\left[\int_0^t b_s^2 \rd s\right]<\infty
\end{eqnarray}
for all $t\in[0,\infty)$. In that case $\{M_t\}$ is called a
square-integrable martingale, and we have the identity
\begin{eqnarray}
{\mathbb E}\left[(M_t-M_0)^2\right] = {\mathbb E}\left[\int_0^t
b_s^2 \rd s\right].
\end{eqnarray}
More generally, we also have the following relation, valid for
$t\geq s\geq 0$, which we call the {\it conditional Wiener-Ito
isometry}:
\begin{eqnarray}
{\mathbb E}_s\left[ (M_t-M_s)^2\right] = {\mathbb E}_s \left[
\int_s^t b_u^2 \rd u\right] . \label{eq:isometry}
\end{eqnarray}

In certain situation we are presented with an equation of the form
(\ref{eq:stc}), and we are told the distribution of $X_0$ and that
the processes $\{a_t\}$ and $\{b_t\}$ are of the form $a_t=
a(X_t,t)$ and $b_t=b(X_t,t)$, where $a(x,t)$ and $b(x,t)$ are
prescribed functions. In that case we have a \emph{stochastic
differential equation} of the form
\begin{eqnarray}
\rd X_t = a(X_t,t) \rd t + b(X_t,t) \rd W_t, \label{eq:diff}
\end{eqnarray}
with initial condition $X_0$. By a `solution' of the stochastic
differential equation (\ref{eq:diff}) we mean the specification of
the probability space $(\Omega,{\mathcal F},{\mathbb P})$ with
filtration $\{{\mathcal F}_t\}$, together with an $\{{\mathcal
F}_t\}$-adapted Brownian motion and an $\{{\mathcal
F}_t\}$-adapted Ito process $\{X_t\}$ satisfying (\ref{eq:diff})
along with the given initial condition.

The extension of these definitions to situations where $\{X_t\}$
and $\{W_t\}$ are multi-dimensional is straightforward. It is also
appropriate in some circumstances to consider processes defined
over a finite time horizon $t\in[0,T]$, $T<\infty$, for which
straightforward modifications of the relevant definitions can also
be formulated.

\section{Dynamics of the energy process}
\label{sec:3}

Now we are in a position to analyse the dynamics of the
energy-based stochastic Schr\"odinger equation (\ref{eq:1.1}) in
more detail. We shall make the following assumptions concerning
the dynamics of the state vector:
\begin{itemize}
\item[(a)] the state-vector process $\{|\psi_t\rangle\}_{0\leq
t<\infty}$ takes values in a finite-dimensional complex Hilbert
space, and is defined on a probability space $(\Omega,{\mathcal
F},{\mathbb P})$ with filtration $\{{\mathcal F}_t\}_{0\leq
t<\infty}$;
\item[(b)] $\{|\psi_t\rangle\}$ is adapted to
$\{{\mathcal F}_t\}$; and
\item[(c)] $\{|\psi_t\rangle\}$ satisfies the stochastic
differential equation (\ref{eq:1.1}) with the given initial
condition $|\psi_0\rangle$.
\end{itemize}

Under these assumptions it is a straightforward exercise in Ito
calculus (see, for example, \cite{brody1}) to show that $\langle
\psi_t|\psi_t\rangle=1$ for all $t\in[0,\infty)$. One is then led
to the following basic result.

\vspace{0.15cm} \noindent {\bf Proposition 1}. {\em The
Hamiltonian process $\{H_t\}$ is an $\{{\mathcal
F}_t\}$-martingale, and the variance process $\{V_t\}$ is an
$\{{\mathcal F}_t\}$-supermartingale.} \vspace{0.15cm}

Proof. We need to show that $\{H_t\}$ satisfies
\begin{eqnarray}
{\mathbb E}_s[H_t]=H_s,  \label{eq:2.1}
\end{eqnarray}
and that $\{V_t\}$ satisfies
\begin{eqnarray}
{\mathbb E}_s[V_t] \leq V_s , \label{eq:2.11}
\end{eqnarray}
where ${\mathbb E}_t[-]$ denotes conditional expectation with
respect to the $\sigma$-algebra ${\mathcal F}_t$. The validity of
these properties can be established as follows. By an application
of Ito's lemma to (\ref{eq:1.2}) and (\ref{eq:1.5}), we infer that
\begin{eqnarray}
{\rm d}H_t = \sigma V_t {\rm d}W_t,  \label{eq:2.2}
\end{eqnarray}
and that
\begin{eqnarray}
{\rm d}V_t = -\sigma^2 V_t^2 {\rm d}t + \sigma \kappa_t {\rm
d}W_t. \label{eq:2.3}
\end{eqnarray}
The process $\{\kappa_t\}$ defined here by
\begin{eqnarray}
\kappa_t = \frac{\langle{\psi}_t|({\hat H}-H_t)^3
|\psi_t\rangle}{\langle{\psi}_t|\psi_t\rangle} \label{eq:2.4}
\end{eqnarray}
measures the skewness of the energy distribution. Integrating
(\ref{eq:2.2}) and (\ref{eq:2.3}) we deduce that
\begin{eqnarray}
H_t = H_0 + \sigma \int_0^t V_u {\rm d}W_u, \label{eq:2.5}
\end{eqnarray}
and that
\begin{eqnarray}
V_t = V_0 -\sigma^2 \int_0^t V_u^2 {\rm d}u + \sigma \int_0^t
\kappa_u {\rm d}W_u . \label{eq:2.6}
\end{eqnarray}
Then on account of the relation
\begin{eqnarray}
{\mathbb E}_s\left[ \int_0^t b_u {\rm d}W_u \right] = \int_0^s b_u
{\rm d}W_u \label{eq:2.999}
\end{eqnarray}
that holds for the stochastic integral of any $\{{\mathcal
F}_t\}$-adapted process $\{b_t\}$ satisfying
\begin{eqnarray}
{\mathbb E}\left[ \int_0^t b_u^2{\rm d}u\right]<\infty,
\label{eq:sigma}
\end{eqnarray}
we deduce the martingale condition (\ref{eq:2.1}) from
(\ref{eq:2.5}). This follows from the fact that $\{V_t\}$ is
bounded. Similarly, it follows as a consequence of (\ref{eq:2.6}),
and the fact that $\{\kappa_t\}$ is bounded, that
\begin{eqnarray}
{\mathbb E}_s\left[V_t\right] = V_s -\sigma^2 {\mathbb E}_s\left[
\int_s^t V_u^2 {\rm d}u \right], \label{eq:2.66}
\end{eqnarray}
which then implies the supermartingale condition (\ref{eq:2.11}).
\hspace*{\fill} $\diamondsuit$

\section{Convergence of the energy variance}
\label{sec:4}

In the case of the Schr\"odinger equation with a time-independent
Hamiltonian, the energy process defined by (\ref{eq:1.2}) is
constant. This is usually interpreted as the quantum mechanical
expression of an energy conservation principle. The martingale
relation (\ref{eq:2.1}) arising in the case of the energy-based
stochastic Schr\"odinger equation can be viewed as a refinement of
this principle.

The supermartingale property (\ref{eq:2.11}) satisfied by the
variance process is the essence of what is meant by a {\it
reduction} process. In the case of the Schr\"odinger equation with
a time-independent Hamiltonian, the variance of the energy is a
constant of the motion. In other words, not only is the
expectation value of the energy fixed, so is the spread. On the
other hand, the spread of the energy is reduced in the case of the
stochastic dynamics of (\ref{eq:1.1}). In fact, the following
result follows as a consequence of equation (\ref{eq:2.6}).

\vspace{0.15cm} \noindent {\bf Proposition 2}. {\it The asymptotic
value of $\{V_t\}$ is given by:
\begin{eqnarray}
\lim_{t\rightarrow\infty}{\mathbb E}\left[V_t\right] = 0.
\label{eq:2.666}
\end{eqnarray}
Therefore, the variance process is a potential.} \vspace{0.15cm}

Proof. First we note that if $X$ and $Y$ are integrable random
variables, and if $X\leq Y$ almost surely, then ${\mathbb
E}[X]\leq {\mathbb E}[Y]$. It follows thus from the
supermartingale condition (\ref{eq:2.11}) by use of the tower
property that if $t\geq u$ then ${\mathbb E}[V_t]\leq {\mathbb E}
[V_u]$. We note that
\begin{eqnarray}
{\mathbb E}\left[\int_0^t b_u {\rm d}W_u \right]=0
\label{eq:2.6666}
\end{eqnarray}
for any $\{{\mathcal F}_t\}$-adapted process $\{b_t\}$ satisfying
(\ref{eq:sigma}). Since the energy skewness process $\{\kappa_t\}$
is bounded, it therefore follows from (\ref{eq:2.6}) that
\begin{eqnarray}
{\mathbb E}\left[V_t\right] &=& V_0 -\sigma^2 {\mathbb E} \left[
\int_0^t V_u^2 {\rm d}u \right] \nonumber \\ &=& V_0 -\sigma^2
\int_0^t {\mathbb E} \left[ V_u^2 \right] {\rm d}u .
\label{eq:2.66666}
\end{eqnarray}
Here we have used Fubini's theorem to interchange the expectation
and the integration. As a consequence, we have the relation
\begin{eqnarray}
{\mathbb E}\left[V_t\right] \leq V_0 -\sigma^2 \int_0^t \left(
{\mathbb E}[V_u]\right)^2 {\rm d}u , \label{eq:2.666666}
\end{eqnarray}
since ${\mathbb E}[V_t^2]\geq ({\mathbb E}[V_t])^2$, which follows
from Jensen's inequality. Now writing
\begin{eqnarray}
v=\lim_{t\to\infty}{\mathbb E}[V_t],
\end{eqnarray}
let us suppose that $v\neq0$. Because ${\mathbb E}[V_t]$ is a
nonnegative, nonincreasing function of time, we have $v\leq
{\mathbb E}[V_t]$. It follows from (\ref{eq:2.666666}) that
${\mathbb E}[V_t]\leq V_0 - \sigma^2v^2t$, which, if $v\neq 0$,
implies that ${\mathbb E}[V_t]$ vanishes at $t=V_0/\sigma^2 v^2$.
However, this is incompatible with the assumption that $v\neq 0$;
it follows that $v=0$, and thus that $\{V_t\}$ is a potential.
\hspace*{\fill} $\diamondsuit$

The same conclusion can be reached by a slightly different line of
argument. Starting with (\ref{eq:2.666666}), we use the fact that
${\mathbb E}[V_t]\leq {\mathbb E} [V_u]$ for $t\geq u$ to infer
that ${\mathbb E}[V_t] \leq V_0-\sigma^2 t ({\mathbb E}[V_t] )^2$,
which implies, on account of the positivity of ${\mathbb
E}_t[V_t]$, that
\begin{eqnarray}
{\mathbb E}[V_t]\leq\sqrt{\frac{V_0}{\sigma^2 t}},
\end{eqnarray}
and hence the claim of the proposition.

Since $V_t$ is nonnegative, Proposition~2 implies that
$\lim_{t\to\infty} V_t =0$ almost surely, i.e. that reduction to a
state of vanishing energy uncertainty occurs with probability one.

\section{Asymptotic properties of the energy}
\label{sec:5}

From the martingale convergence theorem for square-integrable
martingales (see \S\ref{sec:2}) it follows that there exists a
random variable $H_\infty$ defined by
\begin{eqnarray}
H_\infty = \lim_{t\to\infty}H_t, \label{eq:2.777}
\end{eqnarray}
which represents the terminal value of the energy once the
reduction is complete. Thus if we write
\begin{eqnarray}
H_\infty = H_0 + \sigma \int_0^\infty V_u {\rm d}W_u,
\label{eq:2.7}
\end{eqnarray}
it follows as a consequence of (\ref{eq:2.999}) that
\begin{eqnarray}
H_t = {\mathbb E}_t[H_\infty]. \label{eq:2.8}
\end{eqnarray}
Thus $H_\infty$ has the property that it {\it closes} the
martingale $\{H_t\}$. It then follows from (\ref{eq:2.8}) that the
random variable $H_t$ has the interpretation of being the
${\mathcal F}_t$-conditional expectation of the terminal value of
the energy. In particular, we deduce that $H_0={\mathbb
E}[H_\infty]$, which shows that the expectation value of the
Hamiltonian in the initial state agrees with the expectation of
the terminal value of the energy. This result can be viewed as a
justification for the conventional interpretation of the
expectation value of the Hamiltonian.

A similar result can be established in the case of the variance,
which we now proceed to derive. In particular, writing
(\ref{eq:2.66}) in the form
\begin{eqnarray}
{\mathbb E}_t[V_T] = V_t - \sigma^2 {\mathbb E}_t\left[ \int_t^T
V_u^2 \rd u\right],
\end{eqnarray}
it follows that
\begin{eqnarray}
\lim_{T\to\infty}{\mathbb E}_t\left[ V_T\right]=V_t-\sigma^2
\lim_{T\to\infty}{\mathbb E}_t \left[ \int_t^T V_u^2 \rd u\right].
\end{eqnarray}
Since the variance of the energy is bounded, we can invoke the
conditional form of the bounded convergence theorem to interchange
the order of the limit and the expectation on the left-hand side
of this equation. It follows from the fact that
$\lim_{T\to\infty}V_T=0$ almost surely that
\begin{eqnarray}
V_t = \sigma^2 \lim_{T\to\infty}{\mathbb E}_t \left[ \int_t^T
V_u^2 \rd u\right].
\end{eqnarray}
Now we interchange the order of the limit and the conditional
expectation on the right-hand side of this equation by using the
conditional form of the monotone convergence theorem, and we
deduce that
\begin{eqnarray}
V_t =\sigma^2 {\mathbb E}_t \left[ \int_t^\infty V_u^2 \rd u
\right]. \label{eq:xyz}
\end{eqnarray}
On the other hand, it follows as a consequence of (\ref{eq:2.5})
and (\ref{eq:2.7}) that
\begin{eqnarray}
H_\infty-H_t = \sigma \int_t^\infty V_u \rd W_u,
\end{eqnarray}
and therefore, by use of the conditional Wiener-Ito isometry
(\ref{eq:isometry}), that
\begin{eqnarray}
{\mathbb E}_t\left[ (H_\infty-H_t)^2\right] = \sigma^2 {\mathbb
E}_t \left[ \left(\int_t^\infty V_u \rd W_u \right)^2 \right] =
\sigma^2 {\mathbb E}_t \left[ \int_t^\infty V_u^2 \rd u \right].
\label{eq:xyz2}
\end{eqnarray}
Equating the results (\ref{eq:xyz}) and (\ref{eq:xyz2}) we obtain
the fundamental relation:

\vspace{0.15cm} \noindent {\bf Proposition 3}. {\it Let
$\{|\psi_t\rangle\}$ satisfy {\rm (\ref{eq:1.1})} and write
$H_\infty$ for the asymptotic value of the energy martingale
$\{H_t\}$. Then the squared uncertainty of the energy in the state
$|\psi_t\rangle$ is given by
\begin{eqnarray}
V_t = {\mathbb E}_t\left[(H_\infty-{\mathbb E}_t[H_\infty])^2
\right]. \label{eq:2.10}
\end{eqnarray}
} \vspace{0.15cm}

This relation shows that the random variable $V_t$ has the
interpretation of being the {\it conditional variance} of the {\it
terminal} value of the energy. In particular, Proposition~3
demonstrates that the initial squared energy uncertainty $V_0$
agrees with the variance of the terminal value of the energy. This
fact can be viewed as a justification for the conventional
interpretation of the energy uncertainty.

\section{Asymptotic properties of observables that are compatible
with the energy} \label{sec:6}

We now proceed to derive a rather more general result that
includes Proposition~3 as a special case. Let us suppose ${\hat
G}$ is any observable that commutes with ${\hat H}$, and write
$G_t$ and $V_t^G$ for the mean and variance of ${\hat G}$ with
respect to the random state $|\psi_t\rangle$. Thus $G_t=
\langle{\hat G}\rangle_t$ and $V_t^G=\langle({\hat G}-G_t)^2
\rangle_t$, and by use of Ito's lemma we deduce as a consequence
of (\ref{eq:1.1}) that
\begin{eqnarray}
{\rm d}G_t = \sigma \gamma_t {\rm d}W_t, \label{eq:3.1}
\end{eqnarray}
and that
\begin{eqnarray}
{\rm d}V_t^G = -\sigma^2 \gamma_t^2 {\rm d}t + \sigma \delta_t
{\rm d}W_t, \label{eq:3.2}
\end{eqnarray}
where
\begin{eqnarray}
\gamma_t = \langle({\hat G}-G_t)({\hat H}-H_t)\rangle_t
\label{eq:3.3}
\end{eqnarray}
and
\begin{eqnarray}
\delta_t = \langle({\hat G}-G_t)^2({\hat H}-H_t)\rangle_t .
\label{eq:3.4}
\end{eqnarray}

Now we shall show that $\{G_t\}$ is a martingale, and investigate
the nature of the conditional variance representation admitted by
$\{V_t^G\}$. It follows from (\ref{eq:3.1}) and (\ref{eq:3.2})
that
\begin{eqnarray}
G_t = G_0 + \sigma \int_0^t \gamma_u {\rm d}W_u \label{eq:3.5}
\end{eqnarray}
and that
\begin{eqnarray}
V_t^G = V_0^G -\sigma^2 \int_0^t \gamma_u^2 {\rm d}u + \sigma
\int_0^t \delta_u {\rm d}W_u . \label{eq:3.6}
\end{eqnarray}
Thus, since $\{\gamma_t\}$ and $\{\delta_t\}$ are bounded, we see
that $\{G_t\}$ is an $\{{\mathcal F}_t\}$-martingale and that
$\{V_t^G\}$ is an $\{{\mathcal F}_t\}$-supermartingale. Therefore,
by the martingale convergence theorem (see \S\ref{sec:2}) there
exist random variables $G_\infty$ and $V_\infty^G$ such that
\begin{eqnarray}
G_\infty = G_0 + \sigma \int_0^\infty \gamma_u {\rm d}W_u
\label{eq:3.7}
\end{eqnarray}
and
\begin{eqnarray}
V_\infty^G = V_0^G -\sigma^2 \int_0^\infty \gamma_u^2 {\rm d}u +
\sigma \int_0^\infty \delta_u {\rm d}W_u . \label{eq:3.8}
\end{eqnarray}
Taking the conditional expectation of each side of this equation
with respect to ${\mathcal F}_t$, we deduce that
\begin{eqnarray}
{\mathbb E}_t[V_\infty^G] = V_0^G -\sigma^2 {\mathbb E}_t \left[
\int_0^\infty \gamma_u^2 {\rm d}u\right] + \sigma \int_0^t
\delta_u {\rm d}W_u . \label{eq:3.9}
\end{eqnarray}
Solving (\ref{eq:3.9}) for $V_0^G$ and substituting the result
into (\ref{eq:3.6}) we see that
\begin{eqnarray}
V_t^G &=& {\mathbb E}_t[V_\infty^G] + \sigma^2 {\mathbb E}_t
\left[ \int_t^\infty \gamma_u^2 {\rm d}u\right] \nonumber \\ &=&
{\mathbb E}_t[V_\infty^G] + \sigma^2 {\mathbb E}_t \left[
\left(\int_t^\infty \gamma_u {\rm d}W_u \right)^2\right] ,
\label{eq:3.10}
\end{eqnarray}
by use of the conditional Wiener-Ito isometry. Making use of the
fact that
\begin{eqnarray}
G_\infty = G_t + \sigma \int_t^\infty \gamma_u {\rm d}W_u,
\label{eq:3.11}
\end{eqnarray}
which follows from (\ref{eq:3.5}) and (\ref{eq:3.7}), we obtain
the following fundamental relations governing the dynamics of
$\{G_t\}$ and $\{V_t^G\}$:
\begin{eqnarray}
G_t = {\mathbb E}_t[G_\infty]  \label{eq:3.12}
\end{eqnarray}
and
\begin{eqnarray}
V_t^G = {\mathbb E}_t[V_\infty^G] + {\mathbb E}_t \left[
(G_\infty-G_t)^2 \right] .  \label{eq:3.13}
\end{eqnarray}

Equation (\ref{eq:3.12}) shows that the martingale $\{G_t\}$ {\it
closes}, and hence the relation $G_0={\mathbb E}[G_\infty]$ allows
us to identify the initial expectation value $G_0$ with the
expectation of the result obtained for the random variable
$G_\infty$. Equation (\ref{eq:3.13}) then has a natural
interpretation as a {\sl conditional variance relation}. If the
Hamiltonian has a nondegenerate spectrum, then the terminal state
is necessarily both an eigenstate of ${\hat H}$ and ${\hat G}$,
and $V_\infty^G$ vanishes. On the other hand, if ${\hat H}$ has a
degenerate spectrum, then the terminal value of ${\hat H}$ will
not necessarily be an eigenvalue of ${\hat G}$. In that case, the
random variable $V_\infty^G$ is nonvanishing, and takes the value
\begin{eqnarray}
V_\infty^G = \langle\phi_i|{\hat G}^2|\phi_i\rangle -
\langle\phi_i|{\hat G}| \phi_i\rangle^2 \label{eq:3.14}
\end{eqnarray}
with probability
\begin{eqnarray}
\pi_i =  |\langle\psi_0|\phi_i\rangle|^2 , \label{eq:3.15}
\end{eqnarray}
where $|\phi_i\rangle$ is the normalised L\"uders state\footnote{
See \cite{brody1,isham,luders}. We remark, incidentally, that the
L\"uders state has the following geometrical interpretation. In
the case of a Hilbert space of dimension $n+1$, the corresponding
space of pure states is the complex projective space ${\mathbb
C}{\rm P}^n$. If the Hilbert subspace of state vectors of some
given energy $E_i$ has dimension $k+1$, then the corresponding
space of pure states of that energy is a projective hyperplane
$D^k$ of dimension $k$. The complex conjugate of $D^k$ is a
hyperplane ${\bar D}^{n-k-1}$ of dimension $n-k-1$. Clearly $D^k$
and ${\bar D}^{n-k-1}$ do not intersect. The initial state vector
$|\psi_0 \rangle$ corresponds to a point $\psi_0 \in {\mathbb
C}{\rm P}^n$. Therefore the join of $\psi_0$ and $D^k$ is a
hyperplane $Q^{k+1}$ of dimension $k+1$ which intersects the
hyperplane ${\bar D}^{n-k-1}$ at a single point ${\bar\psi}_i$.
Now take the join of $\psi_0$ and ${\bar\psi}_i$. The resulting
line clearly lies in hyperplane $Q^{k+1}$ and thus hits the
hyperplane $D^k$ at a single point, and this point is the L\"uders
state $\phi_i$. The interpretation of ${\bar\psi}_i$, on the other
hand, is as follows: if a measurement is made to determine simply
whether the energy is $E_i$ or not, then in the event of a
negative result the new state of the system will be the point
${\bar\psi}_i$.} corresponding to the eigenvalue $E_i$ of ${\hat
H}$, given the initial state $|\psi_0\rangle$, as defined in
\S\ref{sec:1}. We recall that when ${\hat H}$ has a degenerate
spectrum, the collapse of the wave function induced by
(\ref{eq:1.1}) necessarily leads to one of the L\"uders states, as
shown, e.g., in \cite{brody1}.

It is interesting to observe that, while the variance process
$\{V_t\}$ associated with the Hamiltonian is a potential, the
variance $\{V_t^G\}$ for ${\hat G}$, although a supermartingale,
is not necessarily a potential unless ${\hat H}$ has a
nondegenerate spectrum. Physically this is because a reduction of
the energy induces a complete reduction of a compatible observable
only if the energy spectrum is nondegenerate.

\section{On the generality of the dynamical equation}
\label{sec:7}

Before embarking on an account of our approach to the solution of
the stochastic differential equation (\ref{eq:1.1}), it will be
useful to set this stochastic equation in the context of a more
general family of possible dynamical laws for the state vector
process $\{|\psi_t\rangle\}$. The idea then will be to see what
specific additional physical assumptions are needed to imply that
the dynamics should take the form (\ref{eq:1.1}).

We shall assume as before that $\{|\psi_t\rangle\}_{0\leq
t<\infty}$ is a continuous stochastic process defined on a fixed
probability space $(\Omega,{\mathcal F},{\mathbb P})$ with
filtration $\{{\mathcal F}_t\}$ taking values in a
finite-dimensional complex Hilbert space. For the dynamics of
$\{|\psi_t\rangle\}$ we write
\begin{eqnarray}
{\rm d}|\psi_t\rangle = {\hat\mu}_t |\psi_t\rangle {\rm d}t +
{\hat\sigma}_t |\psi_t\rangle {\rm d}W_t,  \label{eq:3.01}
\end{eqnarray}
where $\{W_t\}$ is an $\{{\mathcal F}_t\}$-Brownian motion. We
assume that $\{|\psi_t\rangle\}$ is adapted to $\{{\mathcal
F}_t\}$, and that so are the operator-valued processes
$\{{\hat\mu}_t\}$ and $\{{\hat\sigma}_t\}$. We call
$\{{\hat\mu}_t\}$ and $\{{\hat\sigma}_t\}$ the operator-valued
drift and volatility of $\{|\psi_t\rangle\}$.

Our first requirement will be that $\{{\hat\mu}_t\}$ and
$\{{\hat\sigma}_t\}$ must be chosen such that the normalisation of
$|\psi_t\rangle$ is preserved for all $t$. We note that the
conjugate of (\ref{eq:3.01}) is
\begin{eqnarray}
{\rm d}\langle{\psi}_t| = \langle{\psi}_t| {\hat\mu}_t^{\dagger}
{\rm d}t + \langle{\psi}_t|{\hat\sigma}_t^{\dagger} {\rm d} W_t.
\label{eq:3.02}
\end{eqnarray}
By virtue of the Ito product rule we have
\begin{eqnarray}
{\rm d}\langle{\psi}_t|\psi_t\rangle = ({\rm d}\langle
{\psi}_t|)|\psi_t\rangle + \langle{\psi}_t|({\rm d}
|\psi_t\rangle) + ({\rm d}\langle{\psi}_t|) ({\rm d}
|\psi_t\rangle) \label{eq:3.03} ,
\end{eqnarray}
and thus by use of (\ref{eq:3.01}) and (\ref{eq:3.02}) we obtain
\begin{eqnarray}
\frac{{\rm d}\langle{\psi}_t|\psi_t\rangle} {\langle{\psi}_t|
\psi_t\rangle} = \left( \langle {\hat\mu}^{\dagger}_t +
{\hat\mu}_t\rangle_t + \langle {\hat \sigma}_t^{\dagger}
{\hat\sigma}_t\rangle_t \right) {\rm d}t + \langle{\hat
\sigma}_t^{\dagger} +{\hat\sigma}_t\rangle_t {\rm d}W_t ,
\label{eq:3.04}
\end{eqnarray}
where for brevity we use the convenient notation
\begin{eqnarray}
\langle{\hat X}_t\rangle_t= \frac{\langle\psi_t|{\hat X}_t|\psi_t
\rangle}{\langle\psi_t|\psi_t \rangle}
\end{eqnarray}
for the expectation value at time $t$ of any operator process
$\{{\hat X}_t\}$. Therefore, the normalisation condition for
$|\psi_t\rangle$ is ensured if the operators
${\hat\mu}^{\dagger}_t + {\hat\mu}_t+ {\hat\sigma}_t^{\dagger}
{\hat\sigma}_t$ and ${\hat\sigma}_t^{\dagger} +{\hat\sigma}_t$
have vanishing expectation values with respect to
$|\psi_t\rangle$. It is a straightforward exercise to verify that
the most general expressions for the drift ${\hat\mu}$ and the
volatility ${\hat\sigma}$ satisfying these conditions are
\begin{eqnarray}
{\hat\mu}_t = -{\rm i}{\hat H}_t - \half{\hat\sigma}_t^\dagger
{\hat\sigma}_t + {\hat J}_t-\langle{\hat J}_t\rangle_t,
\label{eq:3.05}
\end{eqnarray}
and
\begin{eqnarray}
{\hat\sigma}_t = {\rm i}{\hat K}_t + {\hat L}_t- \langle{\hat L}_t
\rangle_t, \label{eq:3.06}
\end{eqnarray}
where $\{{\hat H}_t\}$, $\{{\hat J}_t\}$, $\{{\hat K}_t\}$, and
$\{{\hat L}_t\}$ are arbitrary Hermitian operator-valued
processes.

It should be evident that the process (\ref{eq:1.1}) is obtained
if (a) we let ${\hat K}_t$ and ${\hat J}_t$ vanish for all $t$,
(b) we let ${\hat L}_t=\frac{1}{2}\sigma{\hat H}_t$ for all $t$,
where $\sigma$ is a parameter, and (c) we assume that $\{{\hat
H}_t\}$ is time independent. Let us investigate therefore the
nature of the additional physical conditions that we need to
impose on the general norm-preserving dynamics in order to ensure
that the energy-based model is obtained in accordance with these
specifications.

The general norm-preserving model contains the four
operator-valued processes $\{{\hat H}_t\}$, $\{{\hat J}_t\}$,
$\{{\hat K}_t\}$, and $\{{\hat L}_t\}$. One can think of these
operators as representing properties of the physical environment
in which the quantum system exists. In general, the environment is
changing in a random time-dependent manner. We shall make the
simplifying assumption of a `stationary' environment so that
$\{{\hat H}_t\}$, $\{{\hat J}_t\}$, $\{{\hat K}_t\}$, and $\{{\hat
L}_t\}$ are now replaced by time-independent operators ${\hat H}$,
${\hat J}$, ${\hat K}$, and ${\hat L}$. Thus our first assumption
is that the environment is in a state of stationary equilibrium.

We give the operator ${\hat H}$ the usual interpretation as
representing the total energy of the system. This is justified by
the fact that if ${\hat K}$, ${\hat J}$ and ${\hat L}$ are set to
zero then the conventional Schr\"odinger equation is recovered.
Next we make an assumption that might be called the `universality
of the Hamiltonian'. This is based on the observation that the
Hamiltonian is the only observable that {\it must} exist as an
element of the dynamics of a quantum system. If our dynamical law
is to be universally applicable to any quantum system, then the
only observable that can enter the discussion is ${\hat H}$, and
thus we must require that ${\hat K}$, ${\hat J}$ and ${\hat L}$
are all functions of the Hamiltonian. Thus, in effect, we are
asking that the system should act as its own environment. It is
with this assumption that an element of nonlinearity enters the
dynamics.

Our final physical requirement is that energy should be conserved
in some suitable sense. Now in ordinary quantum mechanics with a
time-independent Hamiltonian, the expectation value of the
Hamiltonian is a constant of the motion. This relation is usually
interpreted by physicists with a certain looseness of language to
mean `conservation of energy', but what it means really is
conservation of the expectation value of the energy. In a
situation where the state is undergoing random changes, the
expectation value of the energy will also change randomly. We can,
nonetheless, impose a slightly weaker condition of conservation
appropriate to this situation by requiring that the process
$\{H_t\}_{0\leq t<\infty}$ should satisfy the martingale relation
\begin{eqnarray}
{\mathbb E}\left[ H_t|{\mathcal F}_s\right] = H_s
\end{eqnarray}
for $s\leq t$. This relation states that the conditional
expectation of the expectation value of the energy at time $t$,
with respect to the $\sigma$-algebra ${\mathcal F}_s$, is the
expectation value of the energy at time $s$. This is the sense in
which the martingale relation provides a characterisation of the
principle of energy conservation.

It follows therefore that we need to analyse the process
$\{H_t\}$, defined as in (\ref{eq:1.2}), and require that its
drift should vanish. This will ensure that $\{H_t\}$ is a
martingale, and that energy is conserved. In fact, we shall impose
a somewhat stronger condition. Let $f(x)$ denote a bounded
function, and write ${\hat f}=f({\hat H})$. We shall require that
for any such operator ${\hat f}$ the corresponding
expectation-value process $\{f_t\}$ defined by
\begin{eqnarray}
f_t = \frac{\langle{\psi}_t|f({\hat H})|\psi_t\rangle}
{\langle{\psi}_t|\psi_t\rangle} \label{eq:100.2}
\end{eqnarray}
should be an $\{{\mathcal F}_t\}$-martingale. This corresponds to
the requirement that not only is the energy conserved in the sense
discussed above but so is the observable associated with any
function of the energy. With this condition in place we have a
suitably general and robust representation of the principle of
energy conservation. If we take the stochastic differential of
$f_t$ in (\ref{eq:100.2}), then by use of the Ito calculus we find
that
\begin{eqnarray}
{\rm d}f_t = 2\left(\langle {\hat J} \hat{f}\rangle_t - \langle
{\hat J}\rangle_t \langle \hat{f}\rangle_t \right) {\rm d}t +2
\left(\langle {\hat L}\hat{f}\rangle_t - \langle {\hat L}
\rangle_t \langle \hat{f}\rangle_t \right) {\rm d}W_t .
\label{eq:2.111}
\end{eqnarray}
Now the martingale condition on $\{f_t\}$ implies that the drift
of $\{f_t\}$ in (\ref{eq:2.111}) must vanish. In other words, we
require that the covariance of the two operators ${\hat J}$ and
${\hat f}$ should vanish for any choice of the function $f(x)$.
Thus in particular if we set ${\hat f}={\hat J}$ then it follows
that the uncertainty of ${\hat J}$ must vanish in the state
$|\psi_t\rangle$, and hence without loss of generality we may
assume that ${\hat J}$ is a constant multiple of the identity
matrix, and therefore drops out of the dynamics.

Finally we consider the roles of ${\hat K}$ and ${\hat L}$ in the
expression for ${\hat \sigma}_t$ in (\ref{eq:3.6}). To this end we
shall examine the dynamics of the squared uncertainty of the
operator ${\hat H}$ in the state $|\psi_t\rangle$. Now so far we
have through our physical considerations specialised the general
dynamics (\ref{eq:3.01}) to the particular case
\begin{eqnarray}
{\rm d}|\psi_t\rangle = \left(-\ri {\hat H}-\half
{\hat\sigma}_t^\dagger {\hat\sigma}_t \right)|\psi_t\rangle {\rd}t
+ {\hat\sigma}_t |\psi_t \rangle {\rd}W_t, \label{eq:3.101}
\end{eqnarray}
where
\begin{eqnarray}
{\hat\sigma}_t = {\rm i}{\hat K} + {\hat L}-\langle{\hat L}
\rangle_t,
\end{eqnarray}
and ${\hat H}$, ${\hat K}$, and ${\hat L}$ are time-independent
and Hermitian, with the further provision that ${\hat K}$ and
${\hat L}$ are both given by functions of ${\hat H}$. We shall
call (\ref{eq:3.101}) the general stationary energy-based
stochastic Schr\"odinger equation.

Let us therefore investigate the extent to which the general
stationary energy-based dynamics (\ref{eq:3.101}) necessarily
leads to state reduction. Writing
\begin{eqnarray}
V_t=\langle{\hat H}^2\rangle_t-\langle{\hat H}\rangle_t^2
\end{eqnarray}
for the variance of ${\hat H}$ with respect to the state
$|\psi_t\rangle$, we obtain
\begin{eqnarray}
\rd V_t=\rd\langle{\hat H}^2\rangle_t-2\langle{\hat H}\rangle_t
\rd\langle{\hat H}\rangle_t-(\rd\langle{\hat H}\rangle_t)^2.
\end{eqnarray}
Now making use of the fact that both $\{\langle{\hat H}^2\rangle_t
\}$ and $\{\langle{\hat H}\rangle_t\}$ are martingales (cf.
\cite{adler}), we immediately infer that $\{V_t\}$ is a
supermartingale. In fact, a calculation gives
\begin{eqnarray}
\fl\hspace{1.0cm} \rd V_t=-4\left\langle ({\hat H}-\langle
H\rangle_t)({\hat L}- \langle L\rangle_t) \right\rangle_t^2 \rd t
+ 2 \left\langle ({\hat H}-\langle H \rangle_t)^2({\hat L}-\langle
L\rangle_t) \right\rangle_t \rd W_t. \label{eq:3.001}
\end{eqnarray}
It is apparent from (\ref{eq:3.001}) that the drift of the energy
variance process is negative in the general stochastic extension
of the Schr\"odinger equation given by (\ref{eq:3.101}). This
demonstrates that {\it the presence of some element of state
reduction or relaxation is a generic feature of the dynamics of
{\rm (\ref{eq:3.101})}, regardless of the specific choice of the
functions determining ${\hat K}$ and ${\hat L}$}.

This result offers some support to the proposal put forward in
Ref.~\cite{adler4} that dynamic reduction in quantum theory might
be an `emergent' phenomenon.

In fact, a short calculation establishes that under the general
energy-based dynamics (\ref{eq:3.101}), the variance process
$\{V_t^L\}$ associated with the operator ${\hat L}$, defined by
$V_t^L=\langle({\hat L}-\langle{\hat L}\rangle_t)^2\rangle_t$,
satisfies the conditions of being a potential, and admits the
following representation as a conditional variance:
\begin{eqnarray}
V_t^L = {\mathbb E}_t[(L_\infty-L_t)^2] ,
\end{eqnarray}
where $L_\infty$ denotes the terminal limiting value of the
martingale $\{L_t\}$ defined by $L_t=\langle{\hat L}\rangle_t$.

Thus, provided the eigenstates of ${\hat L}$ are also eigenstates
of ${\hat H}$, then (\ref{eq:3.101}) necessarily implies a
reduction to energy eigenstates. In what follows we shall
therefore make the simplest choice that ensures this condition,
namely, ${\hat K}=0$ and ${\hat L}=\half\sigma{\hat H}$, where
$\sigma$ is a parameter. Nevertheless, we see that in a general
setting there is scope for some variation in the dynamics of the
state vector from that appearing in (\ref{eq:1.1}). In particular,
we can also consider dropping the stationarity condition. Later in
this paper we present a useful example of a nonstationary
dynamical law.

\section{Conditional probabilities for reduction}
\label{sec:8}

An important special case of the situation described in
\S\ref{sec:6} arises when the observable ${\hat G}$ corresponds to
the projection operator ${\hat\Pi}_i$ onto the subspace of states
with energy $E_i$ (cf. \cite{brody1,adler}). In this case we have
the relations ${\hat H}{\hat\Pi}_i={\hat\Pi}_i{\hat H}$,
${\hat\Pi}_i{\hat\Pi}_j=\delta_{ij}{\hat\Pi}_i$, $\sum_i
{\hat\Pi}_i=1$, and $\sum_i E_i{\hat\Pi}_i={\hat H}$. The spectrum
of ${\hat H}$ may or may not be degenerate.

Writing $\pi_{it}=\langle{\hat\Pi}_i\rangle_t$ for the expectation
value of the operator ${\hat\Pi}_i$ in the state $|\psi_t\rangle$
we deduce as a consequence of the results of \S\ref{sec:6} that
\begin{eqnarray}
\rd\pi_{it}=\sigma\pi_{it}(E_i-H_t)\rd W_t, \label{eq:pi}
\end{eqnarray}
and that
\begin{eqnarray}
\rd v_{it}=-\sigma^2\pi_{it}^2(E_i-H_t)^2\rd t + \sigma\pi_{it}
(1-2\pi_{it})(E_i-H_t)\rd W_t. \label{eq:vi}
\end{eqnarray}
Here $v_{it}$ denotes the variance of the operator ${\hat\Pi}_i$
in the state $|\psi_t\rangle$. We note that in the case of a
projection operator the variance takes the simple form
\begin{eqnarray}
v_{it}=\pi_{it} (1-\pi_{it}). \label{eq:vi2}
\end{eqnarray}
The random variable $\pi_{it}$ has the interpretation of being the
conditional probability that reduction to a state with energy
$E_i$ will occur. In particular, the initial quantities $\pi_i=
\pi_{i0}$ are the Dirac transition probabilities from the initial
state $|\psi_0\rangle$ to a state with energy $E_i$.

It is evident that the process $\{\pi_{it}\}$ is a martingale, and
that this martingale is closed by the random variable
\begin{eqnarray}
\pi_{i\infty}={\bf 1}_{\{H_\infty=E_i\}}. \label{eq:pi1}
\end{eqnarray}
That is to say,
\begin{eqnarray}
\pi_{it}={\mathbb E}_t\left[ \pi_{i\infty}\right] . \label{eq:pi2}
\end{eqnarray}
As a consequence we see that $v_{it}$ can be written in the form
\begin{eqnarray}
v_{it}={\mathbb E}_t\left[ (\pi_{i\infty}-{\mathbb E}_t[
\pi_{i\infty}])^2\right] . \label{eq:vi3}
\end{eqnarray}
In other words, $v_{it}$ can be interpreted as the conditional
variance of the indicator function for collapse to a state of
energy $E_i$. Equation (\ref{eq:vi3}) follows immediately from
(\ref{eq:vi2}) if we make use of the fact that the terminal
indicator function for the energy $E_i$ satisfies $(\pi_{i\infty}
)^2=\pi_{i\infty}$. We see therefore that $\{v_{it}\}$ is a
potential.

Now we proceed to derive another expression for $\{\pi_{it}\}$
that will play a key role in the developments that follow. It is
well known from the theory of stochastic differential equations
that an equation of the form (\ref{eq:pi}) can be integrated. If a
positive process $\{X_t\}$ satisfies an equation of the form
\begin{eqnarray}
\rd X_t=\alpha_t X_t \rd W_t, \label{eq:zzz1}
\end{eqnarray}
and if $\int_0^t \alpha_s^2 \rd s<\infty$ almost surely for all
$t\in[0,\infty)$, then we can write
\begin{eqnarray}
X_t=X_0\exp\left(\int_0^t \alpha_u \rd W_u-\half\int_0^t
\alpha_u^2 \rd u\right), \label{eq:zzz2}
\end{eqnarray}
where $X_0$ is the initial value of the process. If $\{\alpha_t\}$
itself depends in some way on $\{X_t\}$ then one cannot say that
(\ref{eq:zzz2}) `solves' (\ref{eq:zzz1}). In that situation
(\ref{eq:zzz2}) should be regarded as an integral representation
of the stochastic diferential equation (\ref{eq:zzz1}).
Nevertheless we may be able to extract useful information about
the process $\{X_t\}$ by expressing it in this form. In the
present case we can integrate (\ref{eq:pi}) to obtain
\begin{eqnarray}
\pi_{it} =\pi_i \exp\left(\sigma\int_0^t(E_i-H_u)\rd W_u -\half
\sigma^2\int_0^t(E_i-H_u)^2\rd u\right).
\end{eqnarray}
After some straightforward algebraic rearrangement this can be put
in the form
\begin{eqnarray}
\pi_{it} =\frac{\pi_i \exp\left[\sigma E_i\left(W_t+\sigma
\int_0^t H_u\rd u\right) -\frac{1}{2} \sigma^2 E_i^2 t\right]}
{\exp\left(\sigma\int_0^t H_u\rd W_u + \frac{1}{2}\sigma^2\int_0^t
H_u^2 \rd u\right)}. \label{eq:xxx1}
\end{eqnarray}

A further simplification is then achieved if we introduce the
$\{{\mathcal F}_t\}$-adapted process $\{\xi_t\}_{0\leq t< \infty}$
defined by the relation
\begin{eqnarray}
\xi_t=\sigma\int_0^t H_u \rd u + W_t. \label{eq:91}
\end{eqnarray}
The process $\{\xi_t\}$ is evidently a Brownian motion with drift.
Making use of the relation
\begin{eqnarray}
\rd \xi_t = \sigma H_t\rd t + \rd W_t,
\end{eqnarray}
we can then put (\ref{eq:xxx1}) in the form
\begin{eqnarray}
\pi_{it} =\frac{\pi_i \exp\left(\sigma E_i\xi_t -\frac{1}{2}
\sigma^2 E_i^2 t\right)}{\exp\left(\sigma\int_0^t H_s\rd \xi_s -
\frac{1}{2}\sigma^2\int_0^t H_s^2 \rd s\right)}. \label{eq:xxx2}
\end{eqnarray}
Finally, we note that since $\sum_i\pi_{it}=1$, equation
(\ref{eq:xxx2}) leads us to the following identity:
\begin{eqnarray}
\exp\left(\sigma\int_0^t H_u\rd \xi_u - \half\int_0^t H_u^2 \rd
u\right) = \sum_i \pi_i \exp\left(\sigma E_i\xi_t -\half \sigma^2
E_i^2 t\right) . \label{eq:xxx3}
\end{eqnarray}
Inserting this relation in (\ref{eq:xxx2}) we obtain the following
result.

\vspace{0.15cm} \noindent {\bf Proposition 4}. {\it The
conditional probability process $\{\pi_{it}\}$ for reduction to a
state of energy $E_i$ takes the form
\begin{eqnarray}
\pi_{it} =\frac{\pi_i \exp\left(\sigma E_i\xi_t -\frac{1}{2}
\sigma^2 E_i^2 t\right)}{\sum_i \pi_i \exp\left(\sigma E_i\xi_t
-\frac{1}{2} \sigma^2 E_i^2 t\right)}, \label{eq:xxx4}
\end{eqnarray}
where $\xi_t=W_t+\sigma\int_0^t H_u \rd u$. The energy expectation
process $\{H_t\}$ is given by
\begin{eqnarray}
H_{t} =\frac{\sum_i \pi_i E_i \exp\left(\sigma E_i\xi_t -
\frac{1}{2} \sigma^2 E_i^2 t\right)}{\sum_i \pi_i \exp \left(
\sigma E_i\xi_t -\frac{1}{2} \sigma^2 E_i^2 t\right)},
\label{eq:xxx5}
\end{eqnarray}
and the energy variance process $\{V_t\}$ is given by
\begin{eqnarray}
V_{t} =\frac{\sum_i \pi_i (E_i-H_t)^2 \exp\left(\sigma E_i\xi_t -
\frac{1}{2} \sigma^2 E_i^2 t\right)}{\sum_i \pi_i \exp \left(
\sigma E_i\xi_t -\frac{1}{2} \sigma^2 E_i^2 t\right)}.
\label{eq:xxx6}
\end{eqnarray}
} \vspace{0.15cm}

We observe, incidentally, that it suffices to specify the value of
$\xi_t$ to determine $\pi_{it}$, $H_t$, and $V_t$. In other words,
the random behaviour of these quantities is specified entirely
through their dependence on $\xi_t$.

\section{Information theoretic interpretation of the reduction
process} \label{sec:9}

Given the conditional probability $\pi_{it}$ for reduction to an
energy eigenstate with energy $E_i$, we can consider the
associated information entropy $S_t$. Since the conditional
probability approaches the indicator function (\ref{eq:pi1})
asymptotically, we expect the associated entropy to decrease on
average. This idea can be put into more precise terms as follows:

\vspace{0.15cm} \noindent {\bf Proposition 5}. {\it The Shannon
entropy process $\{S_t\}$ associated with the conditional
probability process $\{\pi_{it}\}$ is a potential.}
\vspace{0.15cm}

The Shannon entropy (or information entropy) associated with
$\pi_{it}$ is defined by the expression
\begin{eqnarray}
S_t = -\sum_{i} \pi_{it} \ln \pi_{it}.
\end{eqnarray}
This entropy is associated in a natural way with the {\it random
density matrix} process defined by the conditional expectation of
the terminal state of the system:
\begin{eqnarray}
{\hat R}_t = {\mathbb E}_t \Big[ |\psi_\infty\rangle \langle
\psi_\infty | \Big].
\end{eqnarray}
For each value of $t$, clearly ${\hat R}_t$ is positive
semi-definite and has unit trace. It should also be evident that
\begin{eqnarray}
S_t = -\tr({\hat R}_t\ln{\hat R}_t).
\end{eqnarray}
We note that the process $\{{\hat R}_t\}$ is distinct from the
process $\{{\hat\rho}_t\}$ defined by
\begin{eqnarray}
{\hat \rho}_t = {\mathbb E} \Big[ |\psi_t\rangle\langle\psi_t|
\Big],
\end{eqnarray}
which is deterministic in $t$. Thus the state ${\hat R}_t$
represents the best conditional estimate of the terminal state of
the system, whereas the state ${\hat\rho}_t$ represents the
initial unconditional expectation of the state that the system
will be in at time $t$. Evidently we have ${\hat R}_0=
{\hat\rho}_\infty$. For clarity let us call ${\hat\rho}_t$ the
\emph{von Neumann state}, and ${\hat R}_t$ the \emph{Shannon
state}.

Now if the initial state of the system is a pure state
$|\psi_0\rangle$, with minimum von Neumann entropy, then as the
reduction proceeds the von Neumann state evolves into a mixed
state ${\hat\rho}_t$ with higher entropy. Therefore, the von
Neumann entropy $-{\tr}({\hat\rho}_t\ln{\hat\rho}_t)$ associated
with the mixed state ${\hat\rho}_t$ increases from zero to the
terminal value $-{\tr}({\hat\rho}_\infty \ln{\hat\rho}_\infty)
=-\sum_i\pi_i\ln \pi_i$. On the other hand, the entropy of the
initial Shannon state ${\hat R}_0$ is $-\sum_i\pi_i \ln\pi_i$, and
the entropy of the terminal Shannon state ${\hat R}_\infty$ is
zero.

Thus the evolution $\{{\hat\rho}_t\}$ of the von Neumann state
describes the increase in ignorance that results in the
statistical description of the system as time moves forward;
whereas the evolution $\{{\hat R}_t\}$ of the Shannon state
describes the increase in information that results as the
measurement outcome is revealed. To put the matter another way,
the entropy $S_t$ associated with the Shannon state ${\hat R}_t$
is the negative of the information content generated by the
information flow $\{{\mathcal F}_s\}_{0\leq s\leq t}$ up to time
$t$. In particular, we expect the Shannon entropy $S_t$ to
decrease on average, because more information is gained as the
collapse process progresses. Finally, when the state has reached
an eigenstate, ${\hat R}_\infty$ becomes a pure state. The
proposition above, which we now proceed to prove, asserts that
this is indeed the case.

Proof of Proposition~5. To begin, we determine the dynamical
equation satisfied by the entropy process $\{S_t\}$. We note that,
as a consequence of the dynamical equation (\ref{eq:pi}) for the
conditional probability process, and the use of Ito's lemma, we
have
\begin{eqnarray}
\rd(\ln\pi_{it}) = -\half \sigma^2(E_i-H_t)^2\rd t +
\sigma(E_i-H_t)\rd W_t.
\end{eqnarray}
It follows, by another application of Ito's lemma, that
\begin{eqnarray}
\rd S_t = -\half\sigma^2 V_t\rd t - \sigma\left( \sum_i E_i
\pi_{it}\ln\pi_{it}-H_tS_t\right)\rd W_t, \label{eq:ent}
\end{eqnarray}
where $\{H_t\}$ is the energy process and $\{V_t\}$ is the energy
variance process. We observe that the volatility of $\{S_t\}$ is
the covariance of the energy and the logarithm of the conditional
probability. Since the drift of $\{S_t\}$ is strictly negative we
see that the entropy process is a supermartingale.

To show that $\{S_t\}$ is a potential we need to show that
$\lim_{t\to\infty}{\mathbb E}[S_t]$ vanishes. Because the
conditional probabilities $\{\pi_{it}\}$ are bounded in the range
$0 \leq\pi_{it}\leq1$, the entropy is positive, and is also
bounded, and thus $\lim_{t\to\infty}{\mathbb E}[S_t] = {\mathbb
E}\left[ \lim_{t\to\infty} S_t\right]$ by virtue of the bounded
convergence theorem. On the other hand, (\ref{eq:pi1}) implies
that $\pi_{i\infty}$ is unity if the terminal energy is $E_i$ and
zero otherwise. Therefore, $\lim_{t\to\infty}S_t=0$ almost surely,
and that establishes the result. \hspace*{\fill} $\diamondsuit$

The fact that $\{S_t\}$ is a potential leads to the following
observation concerning the Shannon entropy and energy fluctuations
during the reduction process.

\vspace{0.15cm} \noindent {\bf Proposition 6}. {\it The Shannon
entropy process $\{S_t\}$ is given by the conditional expectation
of the integrated future energy fluctuation level:
\begin{eqnarray}
S_t = \half \sigma^2 {\mathbb E}_t\left[ \int_t^\infty V_s\rd s
\right] . \label{eq:var0}
\end{eqnarray}
} \vspace{0.15cm}

Proof. To derive this result we integrate the dynamical equation
(\ref{eq:ent}) satisfied by the entropy to deduce that
\begin{eqnarray}
S_T = S_0 - \half\sigma^2\int_0^T V_s\rd s - \sigma \int_0^T
\left( \sum_i E_i \pi_{is}\ln\pi_{is}-H_sS_s\right) \rd W_s.
\end{eqnarray}
Taking the conditional expectation of this relation we infer,
after some rearrangement of terms, that
\begin{eqnarray}
{\mathbb E}_t[S_T] = S_t - \half \sigma^2{\mathbb E}_t\left[
\int_t^T V_s \rd s \right] . \label{eq:qwe}
\end{eqnarray}
The identity (\ref{eq:var0}) then follows from the fact that
$\{S_t\}$ is a potential. \hspace*{\fill} $\diamondsuit$

It is interesting to note, incidentally, that if we let $t\to0$ in
(\ref{eq:qwe}), we obtain the following formula for the cumulative
energy fluctuation during the collapse process:
\begin{eqnarray}
\half\sigma^2 \int_0^\infty {\mathbb E}\left[V_s\right] \rd s =
S_0 . \label{eq:var}
\end{eqnarray}
If the information entropy associated with the initial transition
probabilities $\{\pi_i\}$ is large, so that the initial pure state
$|\psi_0\rangle$ is a highly homogenised superposition of the
energy eigenstates, one would expect the energy fluctuations
during the reduction process to be large. Conversely, if this
entropy is small, so that the initial state is close to one or a
few of the eigenstates, then the energy fluctuations during the
reduction process should be small. Proposition~6 makes this
intuition precise. In particular, the right-hand side of
(\ref{eq:var}) measures the entropic uncertainty of the initial
energy dispersion (see, e.g., \cite{deutsch}), and is independent
of the energy spectrum of the system.


\section{Remarks on the ancillary linear dynamics for the state
vector} \label{sec:10}

One of the main goals of this paper is to present in detail a
general method for obtaining the solution to the dynamical
equation (\ref{eq:1.1}). Before embarking upon this, however, we
shall first consider the properties of the linear stochastic
differential equation
\begin{eqnarray}
{\rm d}|\phi_t\rangle = -{\rm i}{\hat H} |\phi_t\rangle {\rm d}t -
\octa \sigma^2{\hat H}^2 |\phi_t\rangle {\rm d}t + \half \sigma
{\hat H}|\phi_t\rangle {\rm d}\xi_t , \label{eq:1.1-1}
\end{eqnarray}
and study the relation of this equation to (\ref{eq:1.1}). The
stochastic differential equation (\ref{eq:1.1-1}), which we shall
call the ancillary equation, plays an important role in the
analysis of (\ref{eq:1.1}). In this section we shall also
introduce some change-of-measure formulae that will be applied in
later sections of the paper.

In the analysis that follows in this section, the process
$\{|\phi_t\rangle\}_{0\leq t<\infty}$ is to be understood as
defined on a fixed probability space $(\Omega, {\mathcal F},
{\mathbb Q})$ with filtration $\{{\mathcal F}_t \}_{0\leq
t<\infty}$ with respect to which $\{\xi_t\}_{0\leq t<\infty}$ is a
standard Brownian motion. The precise relation of the measure
${\mathbb Q}$ appearing here to the measure ${\mathbb P}$
introduced earlier will be specified shortly, as will the relation
between the processes $\{\xi_t\}$ and $\{W_t\}$. In particular,
the process $\{\xi_t\}$ introduced in this section has no \emph{a
priori} relation to the process $\{\xi_t\}$ with the same name
introduced in \S\ref{sec:8}, though in what follows the connection
between these processes is made precise.

The solution to the ancillary equation (\ref{eq:1.1-1}) is given
by
\begin{eqnarray}
|\phi_t\rangle = {\rm e}^{-{\rm i}{\hat H}t + \frac{1}{2} \sigma
{\hat H}\xi_t - \frac{1}{4} \sigma^2{\hat H}^2 t}|\phi_0\rangle ,
\label{eq:1.1-2}
\end{eqnarray}
where $|\phi_0\rangle$ is a prescribed initial state, normalised
to unity. The fact that (\ref{eq:1.1-2}) implies (\ref{eq:1.1-1})
can be verified by a direct application of Ito's lemma
\begin{eqnarray}
{\rm d}|\phi_t\rangle =  |{\dot\phi}(\xi_t,t)\rangle {\rm d}t +
|\phi^{\prime}(\xi_t,t)\rangle {\rm d}\xi_t + \half
|\phi^{\prime\prime}(\xi_t,t)\rangle ({\rm d} \xi_t)^2 ,
\label{eq:1.1-3}
\end{eqnarray}
with $|\phi_t\rangle=|\phi(\xi_t,t)\rangle$, where the function
$|\phi(\xi,t)\rangle$ is defined by
\begin{eqnarray}
|\phi(\xi,t)\rangle = {\rm e}^{-{\rm i}{\hat H}t + \frac{1}{2}
\sigma {\hat H}\xi - \frac{1}{4} \sigma^2{\hat H}^2 t}
|\phi_0\rangle .
\end{eqnarray}
The dot and prime in (\ref{eq:1.1-3}) denote differentiation with
respect to $t$ and $\xi$, respectively.

We see that as a consequence of (\ref{eq:1.1-2}) that the squared
norm of $|\phi_t\rangle$ takes the form
\begin{eqnarray}
\langle\phi_t|\phi_t\rangle=\langle\phi_0|{\rm e}^{\sigma{\hat H}
\xi_t - \frac{1}{2}\sigma^2{\hat H}^2 t}|\phi_0\rangle .
\label{eq:1.1-4}
\end{eqnarray}
Now writing ${\hat H}=\sum_iE_i{\hat\Pi}_i$, where ${\hat\Pi}_i$
as before denotes the projection operator onto the Hilbert
subspace for which ${\hat H}$ takes the value $E_i$, we have
\begin{eqnarray}
\langle\phi_t|\phi_t\rangle=\sum_i \pi_i {\rm e}^{\sigma E_i\xi_t
- \frac{1}{2}\sigma^2 E_i^2 t} .
\end{eqnarray}
In other words, $\langle\phi_t|\phi_t\rangle$ can be expressed as
a \emph{weighted sum of geometric Brownian motions}. Here $\pi_i$
as before signifies the Dirac transition probability
\begin{eqnarray}
\pi_i=\frac{\langle\phi_0|{\hat\Pi}_i|\phi_0\rangle}
{\langle\phi_0|\phi_0\rangle}
\end{eqnarray}
from the initial state $|\phi_0\rangle$ to the L\"uders state
$|\phi_i\rangle$ associated with the initial state
$|\phi_0\rangle$ and the eigenvalue $E_i$. It follows immediately
by virtue of the properties of geometric Brownian motion that the
process $\{\langle\phi_t|\phi_t\rangle\}$ is a martingale in the
${\mathbb Q}$-measure, satisfying
\begin{eqnarray}
{\mathbb E}_s^{\mathbb Q}[\langle\phi_t|\phi_t\rangle]
=\langle\phi_s|\phi_s\rangle.
\end{eqnarray}
This result can also be seen to follow directly from
(\ref{eq:1.1-4}), by an application of Ito's lemma, which shows
the squared norm of $|\phi_t\rangle$ satisfies the dynamical
equation
\begin{eqnarray}
\rd \langle\phi_t|\phi_t\rangle=\sigma H_t \langle\phi_t
|\phi_t\rangle \rd\xi_t, \label{eq:sde.1}
\end{eqnarray}
where the process $\{H_t\}$ is defined by
\begin{eqnarray}
H_t =  \frac{\langle\phi_0|{\hat H}{\rm e}^{\sigma {\hat H}\xi_t -
\frac{1}{2}\sigma^2{\hat H}^2 t}|\phi_0\rangle}
{\langle\phi_0|{\rm e}^{\sigma {\hat H}\xi_t - \frac{1}{2}
\sigma^2 {\hat H}^2 t}|\phi_0\rangle}. \label{eq:1.1-5}
\end{eqnarray}
The stochastic differential equation (\ref{eq:sde.1}) can then be
put in an integral form to give a useful alternative expression
for the squared norm:
\begin{eqnarray}
\langle\phi_t|\phi_t\rangle=\exp\left(\sigma\int_0^t H_u {\rm d}
\xi_u - \half \sigma^2 \int_0^t H_u^2 {\rm d}u\right) .
\label{eq:1.1-6}
\end{eqnarray}

Our intention is to show that the process $\{H_t\}$ defined in
(\ref{eq:1.1-5}) can in fact be identified with the energy process
defined in (\ref{eq:1.2}). For this, we consider the dynamics of
the normalised state vector
\begin{eqnarray}
|\psi_t\rangle = \langle\phi_t| \phi_t \rangle^{-1/2}
|\phi_t\rangle.
\end{eqnarray}
If we write $N_t=\langle\phi_t|\phi_t\rangle^{1/2}$ for the
normalisation factor, then by Ito's lemma we obtain
\begin{eqnarray}
{\rm d}N_t^{-1} = \toct \sigma^2 H_t^2 N_t^{-1} {\rm d}t - \half
\sigma H_t N_t^{-1} {\rm d}\xi_t. \label{eq:1.1-7}
\end{eqnarray}
Hence for the dynamics of the normalised state $|\psi_t\rangle=
N_t^{-1}|\phi_t\rangle$ we have
\begin{eqnarray}
{\rm d}|\psi_t\rangle = N_t^{-1}{\rm d}|\phi_t\rangle +({\rm d}
N_t^{-1})|\phi_t\rangle+({\rm d} N_t^{-1})({\rm d}|\phi_t\rangle),
\end{eqnarray}
and thus
\begin{eqnarray}
\fl{\rm d}|\psi_t\rangle=-{\rm i}{\hat H} |\psi_t\rangle {\rm d}t
-\octa \sigma^2 \left( {\hat H}^2 + 2{\hat H}H_t -3H_t^2\right)
|\psi_t\rangle{\rm d}t+\half({\hat H}-H_t) |\psi_t\rangle {\rm
d}\xi_t . \label{eq:1.1-8}
\end{eqnarray}
This expression can be simplified if we introduce a process
$\{W_t\}$ by the relation
\begin{eqnarray}
W_t = \xi_t - \sigma \int_0^t H_u {\rm d}u . \label{eq:1.1-9}
\end{eqnarray}
Then the dynamics for the normalised state vector
$\{|\psi_t\rangle\}$ can be written
\begin{eqnarray}
{\rm d}|\psi_t\rangle &=& -{\rm i}{\hat H} |\psi_t\rangle {\rm d}t
- \octa \sigma^2({\hat H}-H_t)^2 |\psi_t\rangle {\rm d}t + \half
\sigma ({\hat H}-H_t)|\psi_t\rangle {\rm d}W_t , \label{eq:1.1-10}
\end{eqnarray}
which is identical in form to (\ref{eq:1.1}), and this leaves us
with the problem of the interpretation of the process $\{W_t\}$.

Now $\{\xi_t\}$ is by hypothesis a ${\mathbb Q}$-Brownian motion,
so evidently $\{W_t\}$, as defined in (\ref{eq:1.1-9}), is a
${\mathbb Q}$-Brownian motion with drift. We can, however, find a
new measure ${\mathbb P}$ with respect to which $\{W_t\}$ is a
${\mathbb P}$-Brownian motion. The precise statement is as
follows. Let us fix a finite time $T<\infty$. Then the relevant
change-of-measure density ${\mathbb Q}$-martingale
$\{\Phi_t\}_{0\leq t\leq T}$ appropriate for transforming from
${\mathbb Q}$ to ${\mathbb P}$ over the time horizon $t\in[0,T]$
is defined by $\Phi_t=\langle\phi_t|\phi_t \rangle$, or
equivalently (\ref{eq:1.1-6}). Thus, if $A\in{\mathcal F}_T$
denotes any ${\mathcal F}_T$-measurable set, and if ${\mathbb
E}^{\mathbb Q}$ denotes expectation with respect to the measure
${\mathbb Q}$, then we {\it define} the probability of the event
$A$ with respect to the measure ${\mathbb P}$ by the formula
\begin{eqnarray}
{\mathbb P}(A) = {\mathbb E}^{\mathbb Q}\left[ \Phi_T {\bf 1}_A
\right].
\end{eqnarray}
The theorem of Girsanov (\cite{karatzas,protter,revuz}) allows us
to infer that if $\{\xi_t\}$ is a ${\mathbb Q}$-Brownian motion,
then the process $\{W_t\}_{0\leq t\leq T}$ defined by
(\ref{eq:1.1-9}) is a ${\mathbb P}$-Brownian motion over the given
time horizon.

We note, incidentally, that if $\{m_t\}$ is any ${\mathbb
Q}$-martingale, then the process $\{M_t\}_{0\leq t\leq T}$ defined
by $M_t=m_t/ \Phi_t$ is a ${\mathbb P}$-martingale. In particular,
since the process
\begin{eqnarray}
\langle\phi_t|{\hat H}|\phi_t\rangle &=& \langle\phi_0|{\hat H}
\re^{\sigma{\hat H}\xi_t-\frac{1}{2}\sigma^2{\hat H}^2t}
|\phi_0\rangle \nonumber \\ &=& \sum_i \pi_i E_i \re^{\sigma E_i
\xi_t-\frac{1}{2}\sigma^2 E_i^2 t}
\end{eqnarray}
is a ${\mathbb Q}$-martingale (a sum of geometric Brownian motions
is a martingale), it follows that the energy process $\{H_t\}$
defined by
\begin{eqnarray}
H_t = \frac{\langle\phi_t|{\hat H}|\phi_t\rangle}
{\langle\phi_t|\phi_t\rangle} = \frac{\langle\psi_t|{\hat
H}|\psi_t\rangle} {\langle\psi_t|\psi_t\rangle} \label{eq:1.1-11}
\end{eqnarray}
is a ${\mathbb P}$-martingale. Therefore, for any finite time
horizon $[0,T]$ the dynamics of (\ref{eq:1.1}) can be reproduced
by the following procedure. First, we solve the ancillary equation
(\ref{eq:1.1-1}) with the required initial condition. Next, the
solution thus obtained is used to construct the processes
$\{H_t\}_{0\leq t\leq T}$, $\{\Phi_t\}_{0\leq t\leq T}$, and $\{
|\psi_t\rangle\}_{0\leq t\leq T}$. Finally, the change-of-measure
density martingale $\{\Phi_t\}$ is used to change from the
`ancillary' measure ${\mathbb Q}$ to the `physical' measure
${\mathbb P}$, which is used to interpret the statistical
properties of the dynamics of the quantum system.

With this information at hand, we can now present another useful
characterisation of the dynamics of the state vector process. We
begin with (\ref{eq:1.1}) and (\ref{eq:1.2}), and introduce the
process $\{\xi_t\}$ by use of the relation (\ref{eq:1.1-9}). The
probability space $(\Omega,{\mathcal F},{\mathbb P})$ and the
filtration $\{{\mathcal F}_t\}$ are defined, with respect to which
$\{W_t\}$ is a standard Brownian motion. We introduce on this
probability space the state-vector process $\{|\Psi_t\rangle\}$ by
writing
\begin{eqnarray}
|\Psi_t\rangle = \exp\left( -\ri{\hat H}t
-\frac{1}{4t}(\xi_t-\sigma {\hat H}t)^2\right) |\psi_0\rangle .
\label{eq:**}
\end{eqnarray}
Then it should be evident that
\begin{eqnarray}
|\psi_t\rangle = \frac{|\Psi_t\rangle}{\sqrt{\langle\Psi_t|
\Psi_t\rangle}},
\end{eqnarray}
and hence that $|\Psi_t\rangle$ is an unnormalised form of the
state vector $|\psi_t\rangle$. In fact, so is the state vector
$|\phi_t\rangle$, but $|\Psi_t\rangle$ and $|\phi_t\rangle$ have
different norms.

The significance of the process $\{|\Psi_t\rangle\}$ is that this
process is identical (modulo straightforward minor changes in
notation) to the nonunitary evolution introduced and used by
Pearle~\cite{pearle,pearle1} for the formulation and analysis of
collapse models. In particular, equation (2.1) of \cite{pearle1}
is identical to our equation (\ref{eq:**}) above.
Pearle~\cite{pearle1} asserts that (\ref{eq:**}) represents ``the
most transparent formulation of the energy-based collapse model''.
Although (\ref{eq:**}) does indeed represent a formulation of the
model, it can hardly be regarded as transparent. The problem is
that the definition of $\{\xi_t\}$ involves $|\Psi_t\rangle$, and
hence (\ref{eq:**}) is, in effect, \emph{no more than an integral
representation of the nonlinear stochastic differential equation}
(\ref{eq:1.1}). To put the matter differently, whereas
$\{|\Psi_t\rangle\}$ depends on $\{\xi_t\}$, the probability law
of the process $\{\xi_t\}$ depends on $\{\langle\Psi_t|
\Psi_t\rangle\}$; this is the content of equation (2.2) of
\cite{pearle1}. Thus, when in what follows we speak of obtaining a
`solution' to (\ref{eq:1.1}), it should be emphasised that we are
not merely seeking a `reformulation' such as that represented by
(\ref{eq:**}), or a change-of-measure induced linearisation.

\section{Observation of the energy in the presence of noise}
\label{sec:11}

We now present a general method for obtaining an explicit solution
to the stochastic differential equation (\ref{eq:1.1}). The method
we propose ties in very suggestively with the theory of nonlinear
filtering as developed for example in \cite{shiryaev}.

Let the probability space $(\Omega,{\mathcal F},{\mathbb P})$ be
given, and let $\{{\mathcal G}_t\}$ be a filtration of ${\mathcal
F}$ with respect to which a standard Brownian motion $\{B_t\}$ is
specified, together with an independent random variable $H$. We
assume that $H$ is ${\mathcal G}_0$-measurable, and that it takes
the values $\{E_i\}_{i=1,2,\cdots,N}$ with the probabilities
$\{\pi_i\}_{i=1,2,\cdots,N}$. If we look ahead briefly to the
results that will eventually follow, the random variable $H$ will
have the interpretation of representing the terminal value of the
energy after state reduction, given the Hamiltonian ${\hat H}$ and
the initial state $|\psi_0\rangle$. However, for the moment we
assign no {\it a priori} physical significance to $H$ and
$\{B_t\}$, which are introduced as an ansatz for obtaining a
solution for (\ref{eq:1.1}).

Now suppose we define a random process $\{\xi_t\}_{0\leq t<
\infty}$ according to the scheme
\begin{eqnarray}
\xi_t = \sigma H t + B_t, \label{eq:4.1}
\end{eqnarray}
where $\sigma$ is a positive constant. Since our units are such
that $\hbar=1$, the random variable $H$ can be thought of as
having units of $[{\rm T}]^{-1}$, and hence $\sigma$, $B_t$, and
$\xi_t$ all have units of $[{\rm T}]^{\frac{1}{2}}$. Later we
shall identify $\sigma$ with the parameter appearing in the
dynamical equation (\ref{eq:1.1}), but for the moment we leave its
value unspecified.

The process $\{\xi_t\}$ introduced here has no \emph{a priori}
connection with the process having the same name introduced in
\S\ref{sec:8}. Nevertheless, as we proceed it will be indicated in
what sense these processes can be identified with one another. In
the probability measure ${\mathbb P}$, the process $\{\xi_t\}$
defined by (\ref{eq:4.1}) is a Brownian motion with a random drift
rate $\sigma H$. For each value of $t$ one can think of $\xi_t$ as
providing noisy information about the random variable $H$. That is
to say, given the value of $\xi_t$ one can try to infer
information about the value of $H$. The presence of the
independent noise $B_t$ interferes with this process. In
particular, for small values of $t$, say those for which
$t\ll\sigma^2$, it is typically the case that $|B_t|/\sigma
t\gg1/\sigma^2$. This follows from the fact that ${\mathbb
E}[|B_t|]=\sqrt{2t/\pi}$. Thus if $t\ll\sigma^2$ and if
$|H|\ll1/\sigma^2$, then knowledge of $\xi_t/\sigma t$ provides
little information about the value of $H$. On the other hand, for
large values of $t$ we have $\xi_t/\sigma t\approx H$. We
emphasise that at this point in our analysis the interpretation of
$\{\xi_t\}$ is irrelevant, since it is being introduced as an
ansatz for obtaining the solution to (\ref{eq:1.1}). Nevertheless,
it will be worthwhile to remark as we proceed on various aspects
of the nature of the `information process' $\{\xi_t\}$.

Let $\{{\mathcal F}_t^\xi\}$ denote the filtration generated by
$\{\xi_t\}$. We consider the process $\{H_t\}_{0\leq t\leq\infty}$
generated by the conditional expectation
\begin{eqnarray}
H_t = {\mathbb E}\left[ H|{\mathcal F}_t^\xi\right] .
\label{eq:4.2}
\end{eqnarray}
Intuitively, conditioning with respect to the $\sigma$-algebra
${\mathcal F}_t^\xi$ means conditioning with respect to the
outcome of the random trajectory $\{\xi_s\}_{0\leq s\leq t}$.
Clearly, ${\mathcal F}_t^\xi\subset{\mathcal G}_t$ since knowledge
of $H$ together with $\{B_s\}_{0\leq s\leq t}$ implies knowledge
of $\{\xi_s\}_{0\leq s\leq t}$, although the converse is not the
case.

\vspace{0.15cm} \noindent {\bf Proposition 7}. {\it The
conditional expectation ${\mathbb E} [H|{\mathcal F}_t^\xi]$
represents the best estimate for the value of $H$ given the
trajectory of the process $\{\xi_s\}_{0\leq s\leq t}$ from time
$0$ up to time $t$. }  \vspace{0.15cm}

Proof. Consider the problem of finding an ${\mathcal
F}_t^\xi$-measurable random variable $Y_t$ that minimises the
expected value of the squared deviation of $H$ from $Y_t$, given
the information ${\mathcal F}_t^\xi$. Thus we wish to find a
choice of $Y_t$ that for each $\omega\in\Omega$ minimises
\begin{eqnarray}
J_t = {\mathbb E}[(H-Y_t)^2|{\mathcal F}_t^\xi]. \label{eq:***}
\end{eqnarray}
Since $Y_t$ is assumed to be ${\mathcal F}_t^\xi$-measurable, we
have
\begin{eqnarray}
{\mathbb E}[(H-Y_t)^2|{\mathcal F}_t^\xi] = {\mathbb
E}[H^2|{\mathcal F}_t^\xi] -2 Y_t {\mathbb E}[H|{\mathcal
F}_t^\xi] + Y_t^2 .
\end{eqnarray}
Now setting $Y_t={\mathbb E}[H|{\mathcal F}_t^\xi]+Z_t$, where
$Z_t$ is any ${\mathcal F}_t^\xi$-measurable random variable, we
find that
\begin{eqnarray}
J_t = {\mathbb E}\left[(H-H_t)^2|{\mathcal F}_t^\xi\right]+Z_t^2,
\label{eq:*****}
\end{eqnarray}
where $H_t={\mathbb E}[H|{\mathcal F}_t^\xi]$. Therefore, $J_t$
achieves its minimum if and only if $Z_t=0$. \hspace*{\fill}
$\diamondsuit$

The intuition behind this result is as follows. We can think of
$H$ as being a hidden variable. Its value is hidden by virtue of
the noise process $\{B_t\}$. The best estimate available at time
$t$ for the value of $H$ is the process $\{H_t\}$ defined by
(\ref{eq:4.2}). Our goal now is to show that $\{H_t\}$ can be
identified with the energy expectation process (\ref{eq:1.2})
associated with the standard energy-based stochastic extension of
the Schr\"odinger equation.

\section{Optimal estimation of the energy}
\label{sec:12}

We proceed in this section to calculate the conditional
expectation (\ref{eq:4.2}) to establish the following useful
result.

\vspace{0.15cm} \noindent {\bf Proposition 8}. {\it Let $H$ be a
random variable taking the value $E_i$ with probability $\pi_i$
$(i=1,2,\ldots,n)$, and set $\xi_t=\sigma Ht+B_t$ for $0\leq
t<\infty$, where $\sigma$ is a constant and the Brownian motion
$\{B_t\}$ is independent of $H$. Then the conditional expectation
$H_t = {\mathbb E}[ H|{\mathcal F}_t^\xi]$ is given by
\begin{eqnarray}
H_t = \frac{\sum_{i} \pi_i E_i \exp\left( \sigma E_i \xi_t -\half
\sigma^2 E_i^2 t\right)}{\sum_{i} \pi_i \exp\left( \sigma E_i
\xi_t-\half \sigma^2 E_i^2 t\right)}. \label{eq:4.1400}
\end{eqnarray}
} \vspace{0.15cm}

Proof. First, we observe that $\{\xi_t\}$ is a Markov process. To
establish that $\{\xi_t\}$ is Markovian, we need to show that for
all $T\geq t$ the conditional probability distribution of $\xi_T$
given the history $\{\xi_s\}_{0\leq s\leq t}$ is equal to the
conditional probability distribution of $\xi_T$ given the value
$\xi_t$ of the process at time $t$ alone. In other words, we need
to establish the following result:

\vspace{0.15cm} \noindent {\bf Lemma 1}. {\it Let $\xi_t=\sigma Ht
+B_t$, where $H$ is a random variable taking the values $E_i$
$(i=1,2,\ldots,N)$ with probability ${\mathbb P}(H=E_i)=\pi_i$,
$\sigma$ is a constant, and $\{B_t\}$ is a standard ${\mathbb
P}$-Brownian motion, independent of $H$. Then for all $T\geq t$
and for all $x\in{\mathbb R}$ we have
\begin{eqnarray}
{\mathbb P}( \xi_T\leq x|{\mathcal F}_t^\xi)= {\mathbb P}( \xi_T
\leq x|\xi_t).
\end{eqnarray}
} \vspace{0.15cm}

Proof of Lemma 1. It suffices to show that
\begin{eqnarray}
{\mathbb P}\left( \xi_t\leq x| \xi_s,\xi_{s_1},\xi_{s_2},\ldots,
\xi_{s_k}\right) = {\mathbb P}\left( \xi_t\leq x|\xi_s\right)
\end{eqnarray}
for any collection of times $t,s,s_1,s_2,\ldots,s_k$ such that
$t\geq s\geq s_1\geq s_2\geq\cdots\geq s_k>0$. Now it is a
remarkable property of Brownian motion that for any times $t,s,
s_1$ satisfying $t>s>s_1>0$ one can show that
\begin{eqnarray}
B_t\ {\rm and}\ \frac{B_s}{s}-\frac{B_{s_1}}{s_1}\ {\rm
are~independent}.
\end{eqnarray}
More generally, if $s>s_1>s_2>s_3>0$, we find that
\begin{eqnarray}
\frac{B_s}{s}-\frac{B_{s_1}}{s_1}\ {\rm and}\ \frac{B_{s_2}}{s_2}
-\frac{B_{s_3}}{s_3}\ {\rm are~independent}.
\end{eqnarray}
In each case the result stated follows after a calculation of the
covariance of the indicated variables. Next we note that
\begin{eqnarray}
\frac{\xi_s}{s}-\frac{\xi_{s_1}}{s_1}=\frac{B_s}{s}-
\frac{B_{s_1}}{s_1}.
\end{eqnarray}
It follows therefore that
\begin{eqnarray}
\fl{\mathbb P}\left( \xi_t\leq x| \xi_s,\xi_{s_1}, \xi_{s_2},
\ldots,\xi_{s_k}\right)&=&{\mathbb P}\left(\xi_t\leq x\Big| \xi_s,
\frac{\xi_s}{s}-\frac{\xi_{s_1}}{s_1}, \frac{\xi_{s_1}}{s_1}-
\frac{ \xi_{s_2}}{s_2}, \ldots, \frac{\xi_{s_{k-1}}}{s_{k-1}} -
\frac{\xi_{s_k}}{s_k}\right)\nonumber \\ && \hspace{-3.0cm}
={\mathbb P}\left( \xi_t\leq x\Big| \xi_s, \frac{B_s}{s}-
\frac{B_{s_1}}{s_1}, \frac{B_{s_1}}{s_1}-\frac{B_{s_2}}{s_2},
\ldots, \frac{B_{s_{k-1} }}{s_{k-1}}-\frac{B_{s_k}}{s_k}\right).
\end{eqnarray}
However, since $\xi_t$ and $\xi_s$ are independent of $B_s/s-
B_{s_1}/s_1$, $B_{s_1}/s_1-B_{s_2}/s_2$, $\ldots$, $B_{s_{k-1}}
/s_{k-1}-B_{s_k}/s_k$, the desired result of Lemma~1 follows.
\hspace*{\fill} $\diamondsuit$

Continuing with the proof of Proposition 8, we note next that
\begin{eqnarray}
{\mathbb E}\left[ H|{\mathcal F}_t^\xi\right] = {\mathbb E}\left[
H|\xi_t\right]. \label{eq:4.2-2}
\end{eqnarray}
That is to say, rather than conditioning on the
$\sigma$-subalgebra ${\mathcal F}_t^\xi$ generated by
$\{\xi_s\}_{0\leq s\leq t}$ it suffices to condition on $\xi_t$
alone (conditioning with respect to a random variable means
conditioning with respect to the $\sigma$-algebra generated by
that random variable). The additional information in
$\{\xi_s\}_{0\leq s\leq t}$ does not allow us to improve the
estimate of $H$ once $\xi_t$ has been given. This follows from the
fact that $\{\xi_t\}$ is Markovian and that
\begin{eqnarray}
\lim_{t\rightarrow\infty}\frac{\xi_t}{t} = \sigma H.
\end{eqnarray}

To calculate ${\mathbb E}[H|\xi_t]$, we require a version of the
Bayes formula applicable when we consider the probability of a
discrete random variable conditioned on the value of a continuous
random variable. In this connection we recall for convenience that
for discrete random variables $A$ and $B$ that take on the values
$A=A_i$ ($i=1,2,\cdots,n$) and $B=B_j$ ($j=1,2,\cdots,m$) with
probabilities $q_i$ and $r_j$, respectively, then we have the
classical Bayes formula
\begin{eqnarray}
{\mathbb P}(A=A_i|B=B_j) = \frac{{\mathbb P}(A=A_i){\mathbb P}
(B=B_j|A=A_i)}{{\mathbb P}(B=B_j)},
\end{eqnarray}
or equivalently
\begin{eqnarray}
{\mathbb P}(A=A_i|B=B_j) = \frac{q_i{\mathbb P} (B=B_j|A=A_i)}
{\sum_{i=1}^n q_i {\mathbb P}(B=B_j|A=A_i)} , \label{eq:4.4}
\end{eqnarray}
since the marginal probability for the random variable $B$ can be
written
\begin{eqnarray}
{\mathbb P}(B=B_j) = \sum_{i=1}^n q_i {\mathbb P}(B=B_j|A=A_i) .
\label{eq:4.5}
\end{eqnarray}

Alternatively, instead of conditioning directly with respect to
the event $B=B_j$ we can condition with respect to the random
variable $B$, and write
\begin{eqnarray}
{\mathbb P}(A=A_i|B) &=& \frac{{\mathbb P}(A=A_i){\mathbb P} (B|A
=A_i)}{{\mathbb P}(B)} \nonumber \\ &=& \frac{q_i{\mathbb P} (B
|A=A_i)} {\sum_{i=1}^n q_i {\mathbb P}(B|A=A_i)}, \label{eq:4.6}
\end{eqnarray}
where ${\mathbb P}(B)$ is the random variable that takes the value
$r_j={\mathbb P}(B=B_j)$ when $B$ takes value $B_j$, and ${\mathbb
P}(B|A=A_i)$ is the random variable that takes the value ${\mathbb
P}(B=B_j|A=A_i)$ when $B$ takes value $B_j$. Clearly,
\begin{eqnarray}
{\mathbb P}(B) = \sum_{i=1}^n q_i {\mathbb P}(B|A=A_i) .
\label{eq:4.7}
\end{eqnarray}

For our purpose we need the analogue of (\ref{eq:4.6}) applicable
in the situation where $A$ is a discrete random variable and $B$
is a continuous random variable. In that case
\begin{eqnarray}
{\mathbb P}(A=A_i|B) = \frac{{\mathbb P}(A=A_i)\rho(B| A=A_i)}
{\rho(B)},
\end{eqnarray}
or equivalently
\begin{eqnarray}
{\mathbb P}(A=A_i|B) = \frac{q_i\rho(B|A=A_i)} {\sum_{i=1}^n q_i
\rho(B|A=A_i)} , \label{eq:4.8}
\end{eqnarray}
since
\begin{eqnarray}
\rho(B) = \sum_{i=1}^n q_i \rho(B|A=A_i) . \label{eq:4.9}
\end{eqnarray}
Here $\rho(x)$ denotes the density function of the continuous
random variable $B$, so
\begin{eqnarray}
{\mathbb P}(B<b) = \int_{-\infty}^b \rho(x) {\rm d}x ,
\label{eq:4.10}
\end{eqnarray}
and $\rho(x|A=A_i)$ is the conditional density of $B$ given
$A=A_i$, so
\begin{eqnarray}
{\mathbb P}(B<b|A=A_i) = \int_{-\infty}^b \rho(x|A=A_i) {\rm d}x .
\label{eq:4.11}
\end{eqnarray}
The random variable $\rho(B)$, {\it resp}. $\rho(B|A=A_i)$, takes
the value $\rho(b)$, {\it resp}. $\rho(b|A=A_i)$, when $B$ takes
the value $b$.

Equation (\ref{eq:4.8}) is the version of the Bayes formula we
require in order to determine the conditional expectation
(\ref{eq:4.2}). In particular, since $\xi_t$ is a continuous
random variable, we have
\begin{eqnarray}
{\mathbb P}(H=E_i|\xi_t) &=&  \frac{{\mathbb P}(H=E_i)\rho
(\xi_t|H=E_i)}{\rho(\xi_t)} \nonumber \\ &=& \frac{{\mathbb
P}(H=E_i)\rho(\xi_t|H=E_i)}{\sum_{i} {\mathbb P}
(H=E_i)\rho(\xi_t|H=E_i)} \nonumber \\ &=& \frac{\pi_i\rho
(\xi_t|H=E_i)}{\sum_{i} \pi_i \rho(\xi_t|H=E_i)}. \label{eq:4.12}
\end{eqnarray}
Here $\rho(\xi_t|H=E_i)$ denotes the conditional density function
for the random variable $\xi_t$ given that $H=E_i$. Since
$\{B_t\}$ is a standard Brownian motion in the ${\mathbb
P}$-measure, the conditional probability density for $\xi_t$ is
Gaussian and is given by
\begin{eqnarray}
\rho(\xi_t|H=E_i) = \frac{1}{\sqrt{2\pi t}} \exp\left(-
\frac{1}{2t} (\xi_t-\sigma E_i t)^2\right) . \label{eq:4.13}
\end{eqnarray}
It follows from the Bayes law (\ref{eq:4.12}) that the desired
conditional probability for the random variable $H$ is given by
\begin{eqnarray}
{\mathbb P}(H=E_i|\xi_t) = \frac{\pi_i \exp\left( \sigma E_i \xi_t
-\half \sigma^2 E_i^2 t\right)}{\sum_{i} \pi_i \exp\left( \sigma
E_i \xi_t-\half \sigma^2 E_i^2 t\right)}. \label{eq:4.133}
\end{eqnarray}
Therefore, we deduce that
\begin{eqnarray}
H_t &=& {\mathbb E}[H|\xi_t] \nonumber \\ &=& \sum_{i} E_i
{\mathbb P}(H=E_i|\xi_t) \nonumber \\ &=&  \frac{\sum_{i} \pi_i
E_i \exp\left( \sigma E_i \xi_t-\half \sigma^2 E_i^2
t\right)}{\sum_{i} \pi_i \exp\left( \sigma E_i \xi_t-\half
\sigma^2 E_i^2 t\right)}. \label{eq:4.14}
\end{eqnarray}
That concludes the proof of Proposition 7. \hspace*{\fill}
$\diamondsuit$

More generally (see, e.g., \cite{yor}), a similar argument
establishes that for any bounded function $x\to f(x)$ we have
\begin{eqnarray}
{\mathbb E}\left[ f(H)\Big|{\mathcal F}_t^\xi\right] =
\frac{\sum_{i} \pi_i f(E_i) \exp\left( \sigma E_i \xi_t-\half
\sigma^2 E_i^2 t\right)}{\sum_{i} \pi_i \exp\left( \sigma E_i
\xi_t-\half \sigma^2 E_i^2 t\right)}. \label{eq:4.141}
\end{eqnarray}

\section{Existence of the innovation process}
\label{sec:13}

We now proceed to establish the following basic result.

\vspace{0.15cm} \noindent {\bf Proposition 9}. {\it Let
$\{\xi_t\}$ and $\{H_t\}$ be defined as in Proposition 8. Then the
process $\{W_t\}$ defined by
\begin{eqnarray}
W_t = \xi_t - \sigma \int_0^t H_s {\rm d}s \label{eq:4.17}
\end{eqnarray}
is an $\{{\mathcal F}_t^\xi\}$-Brownian motion.} \vspace{0.15cm}

Proof. Starting with the relation $\xi_t=\sigma Ht+B_t$ we define
$\{W_t\}$ as above, with $\{H_t\}$ defined as in (\ref{eq:4.2}).
To show that $\{W_t\}$ is an $\{{\mathcal F}_t^\xi\}$-Brownian
motion it suffices to show that $\{W_t\}$ is an $\{{\mathcal
F}_t^\xi\}$-martingale and that $(\rd W_t)^2=\rd t$. First, we
shall demonstrate that $\{W_t\}$ is an $\{{\mathcal
F}_t^\xi\}$-martingale. Letting $t\leq T$ we deduce that
\begin{eqnarray}
{\mathbb E}\left[W_T\Big|{\mathcal F}_t^\xi\right]&=& {\mathbb E}
\left[\xi_T\Big| {\mathcal F}_t^\xi\right]-\sigma {\mathbb E}
\left[\int_0^T H_s {\rm d}s\Big| {\mathcal F}_t^\xi\right]
\nonumber \\ &=& \sigma T {\mathbb E}\left[ H\Big| {\mathcal
F}_t^\xi\right] + {\mathbb E}\left[ B_T\Big| {\mathcal
F}_t^\xi\right] -\sigma {\mathbb E} \left[\int_0^T H_s {\rm
d}s\Big| {\mathcal F}_t^\xi\right] \nonumber \\ &=& \sigma T
{\mathbb E}\left[ H\Big| {\mathcal F}_t^\xi\right] + {\mathbb
E}\left[ B_T\Big| {\mathcal F}_t^\xi\right] -\sigma \int_0^T
{\mathbb E} \left[ H_s \Big| {\mathcal F}_t^\xi\right]{\rm d}s,
\label{eq:96}
\end{eqnarray}
by Fubini's theorem. Next, we note that
\begin{eqnarray}
\int_0^T {\mathbb E} \left[ H_s \Big| {\mathcal F}_t^\xi\right]
{\rm d}s &=& \int_0^t {\mathbb E} \left[ H_s \Big| {\mathcal
F}_t^\xi\right]{\rm d}s + \int_t^T {\mathbb E} \left[ H_s \Big|
{\mathcal F}_t^\xi\right]{\rm d}s \nonumber \\ &=& \int_0^t H_s
{\rm d}s + \int_t^T H_t {\rm d}s \nonumber \\ &=& \int_0^t H_s
{\rm d}s + (T-t) H_t . \label{eq:97}
\end{eqnarray}
Here we have used the fact that $\{H_t\}$ is an $\{{\mathcal
F}_t^\xi\}$-martingale. Substituting (\ref{eq:97}) in
(\ref{eq:96}) we obtain
\begin{eqnarray}
{\mathbb E}\left[W_T\Big|{\mathcal F}_t^\xi\right] = \sigma t
{\mathbb E}\left[ H\Big| {\mathcal F}_t^\xi\right] + {\mathbb E}
\left[ B_T\Big| {\mathcal F}_t^\xi\right] - \sigma \int_0^t H_s
{\rm d}s . \label{eq:98}
\end{eqnarray}
Finally, we observe that by the tower property of conditional
expectation we have
\begin{eqnarray}
{\mathbb E}\left[B_T\Big|{\mathcal F}_t^\xi\right] = {\mathbb E}
\left[ {\mathbb E}\left[B_T| {\mathcal F}_t^B,H\right]
\Big|{\mathcal F}_t^\xi \right] = {\mathbb E}\left[B_t\Big|
{\mathcal F}_t^\xi \right] . \label{eq:99}
\end{eqnarray}
Inserting this in (\ref{eq:98}) we obtain
\begin{eqnarray}
{\mathbb E}\left[W_T\Big|{\mathcal F}_t^\xi\right] &=& \sigma t
H_t + {\mathbb E}\left[ B_t\Big| {\mathcal F}_t^\xi\right] -\sigma
\int_0^t H_s {\rm d}s \nonumber \\ &=& {\mathbb E}\left[(\sigma t
H +B_t)\Big| {\mathcal F}_t^\xi\right] -\sigma \int_0^t H_s {\rm
d}s \nonumber \\ &=& {\mathbb E}\left[ \xi_t\Big| {\mathcal
F}_t^\xi \right] -\sigma \int_0^t H_s \rd s = W_t , \label{eq:100}
\end{eqnarray}
and this establishes that $\{W_t\}$ is an $\{{\mathcal
F}_t^\xi\}$-martingale. Next, we observe that since
\begin{eqnarray}
{\rm d}W_t=\sigma (H-H_t){\rm d}t+{\rm d} B_t,
\end{eqnarray}
it follows at once that $({\rm d}W_t)^2={\rm d}t$. Together with
the fact that $\{W_t\}$ is an $\{{\mathcal F}_t^\xi\}$-martingale
we conclude that $\{W_t\}$ is an $\{{\mathcal F}_t^\xi\}$-Brownian
motion. \hspace*{\fill} $\diamondsuit$

We call $\{W_t\}$ the {\sl innovation process} associated with the
dynamics of the wave function. The significance of the fact that
$\{W_t\}$ is an $\{{\mathcal F}_t^\xi\}$-Brownian motion is that
the process $\{\xi_t\}$ as defined in (\ref{eq:4.1}) satisfies a
diffusion equation of the form
\begin{eqnarray}
{\rm d}\xi_t = \sigma H_t {\rm d}t + {\rm d}W_t , \label{eq:4.222}
\end{eqnarray}
where $H_t=H(\xi_t,t)$. As a result, one can prove that ${\mathcal
F}_t^{\xi}={\mathcal F}_t^W$; that is to say, the information set
generated by $\{W_t\}$ is equivalent to that generated by
$\{\xi_t\}$. It is the innovation process $\{W_t\}$, and not the
noise $\{B_t\}$, that `drives' the dynamics of the state vector
process $\{|\psi_t\rangle\}$ in (\ref{eq:1.1}).

Now let $|\psi_0\rangle$ be the initial normalised state vector of
the quantum system, and let ${\hat\Pi}_i$ denote for each value of
$i$ the projection operator onto the subspace of Hilbert space
corresponding to the energy eigenvalue $E_i$, which may be
degenerate. As before, we let
\begin{eqnarray}
|\phi_i\rangle=\pi_i^{-1/2}{\hat\Pi}_i |\psi_0\rangle
\end{eqnarray}
denote the L\"uders state corresponding to $E_i$, and we write
\begin{eqnarray}
\pi_{it} = {\mathbb P}\left( H=E_i|\xi_t\right) \label{eq:4.23}
\end{eqnarray}
for the process defined by (\ref{eq:4.133}).

\vspace{0.15cm} \noindent {\bf Theorem 1}. {\it The solution of
the stochastic differential equation
\begin{eqnarray}
{\rm d}|\psi_t\rangle = -{\rm i}{\hat H} |\psi_t\rangle {\rm d}t -
\octa \sigma^2({\hat H}-H_t)^2 |\psi_t\rangle {\rm d}t + \half
\sigma ({\hat H}-H_t)|\psi_t\rangle {\rm d}W_t \label{eq:1.1-50}
\end{eqnarray}
with initial condition $|\psi_0\rangle$ is given by
\begin{eqnarray}
|\psi_t\rangle = \sum_{i} {\re}^{-{\rm i}E_i t} \pi_{it}^{1/2}
|\phi_i\rangle . \label{eq:4.24}
\end{eqnarray}
Here $|\phi_i\rangle$ denotes the L\"uders state for the
eigenvalue $E_i$, and
\begin{eqnarray}
\pi_{it}= \frac{\pi_i \exp\left( \sigma E_i \xi_t-\half \sigma^2
E_i^2 t\right)}{\sum_{i} \pi_i \exp\left( \sigma E_i \xi_t-\half
\sigma^2 E_i^2 t\right)}, \label{eq:4.245}
\end{eqnarray}
where $\xi_t=\sigma H t+B_t$. The random variable $H$ takes the
value $\{E_i\}$ with the probabilities $\{\pi_i\}$, and $\{B_t\}$
is a Brownian motion independent of $H$. The process $\{H_t\}$ is
defined by $H_t=\sum_i E_i \pi_{it}$, and the $\{{\mathcal
F}_t^\xi\}$-Brownian motion $\{W_t\}$ is given by $W_t=\xi_t-
\sigma \int_0^t H_u\rd u$. } \vspace{0.15cm}

Proof. It is a straightforward exercise to verify that
(\ref{eq:4.24}) satisfies the stochastic differential equation
(\ref{eq:1.1}) with the given initial condition. In particular, by
applying Ito's lemma to (\ref{eq:4.245}) and using the relation
(\ref{eq:4.222}), we can verify that $\{\pi_{it}\}$ satisfies
\begin{eqnarray}
{\rm d}\pi_{it} = \sigma (E_i-H_t)\pi_{it} {\rm d}W_t.
\label{eq:4.25}
\end{eqnarray}
Then with another application of Ito's lemma we deduce that
\begin{eqnarray}
{\rm d}\pi_{it}^{1/2} = - \octa \sigma^2 (E_i-H_t)^2
\pi_{it}^{1/2} {\rm d}t + \half \sigma (E_i-H_t)\pi_{it}^{1/2}
{\rm d} W_t , \label{eq:4.26}
\end{eqnarray}
and with this relation at hand a short calculation shows that
(\ref{eq:4.24}) satisfies (\ref{eq:1.1}). \hspace*{\fill}
$\diamondsuit$

\section{Direct verification of the reductive property}
\label{sec:14}

Thus, summing up, the stochastic equation (\ref{eq:1.1}) can be
solved as follows. We let $H$ be a random variable taking values
$\{E_i\}$ with the probabilities $\{\pi_i\}$ defined by
(\ref{eq:3.15}), or equivalently by
\begin{eqnarray}
\pi_i = \frac{\langle\psi_0|{\hat\Pi}_i|\psi_0\rangle}
{\langle\psi_0|\psi_0\rangle} .
\end{eqnarray}
Letting $\{B_t\}$ denote an independent Brownian motion, we define
the process $\{\xi_t\}_{0\leq t<\infty}$ by writing $\xi_t=\sigma
Ht+B_t$. The solution of (\ref{eq:1.1}) is then given by
(\ref{eq:4.24}) or equivalently
\begin{eqnarray}
|\psi_t\rangle = \frac{\sum_{i} \pi_i^{1/2} \exp\left( -{\rm i}E_i
t+\frac{1}{2}\sigma E_i\xi_t - \frac{1}{4}\sigma^2 E_i^2
t\right)|\phi_i\rangle} {\left( \sum_{i} \pi_i \exp\left( \sigma
E_i\xi_t-\frac{1}{2}\sigma^2 E_i^2 t\right)\right)^{1/2}},
\end{eqnarray}
where the $\{{\mathcal F}_t^\xi\}$-Brownian motion $\{W_t\}$
driving $\{|\psi_t\rangle\}$ in (\ref{eq:1.1-50}) is given by
(\ref{eq:4.17}), with $\{H_t\}$ defined as in (\ref{eq:4.14}).

The fact that (\ref{eq:4.14}) defines a reduction process for the
energy can be verified directly as follows. Suppose, in a
particular realisation of the process $\{H_t\}$, the random
variable $H$ takes the value $E_j$ for some specific choice of the
index $j$. That is to say, we condition on the event $H=E_i$.
Substituting $\xi_t=\sigma E_j t+B_t$ for the corresponding
realisation of $\{H_t\}$, we have
\begin{eqnarray}
H_t &=& \frac{\sum_{i} \pi_i E_i \exp\left( \sigma E_i B_t-\half
\sigma^2 E_i(E_i-2E_j) t\right)}{\sum_{i} \pi_i \exp\left(
\sigma E_i B_t-\half \sigma^2 E_i(E_i-E_j) t\right)} \nonumber \\
&=& \frac{\sum_{i} \pi_i E_i \exp\left( \sigma (E_i-E_j) B_t-\half
\sigma^2 (E_i-E_j)^2 t\right)}{\sum_{i} \pi_i \exp\left( \sigma
(E_i-E_j) B_t-\half \sigma^2 (E_i-E_j)^2 t\right)} \nonumber \\
&=& \frac{\pi_j E_j + \sum_{i(\neq j)} \pi_i E_i \exp\left( \sigma
(E_i-E_j) B_t-\half \sigma^2 (E_i-E_j)^2 t\right)}{\pi_j +
\sum_{i(\neq j)} \pi_i \exp\left( \sigma (E_i-E_j) B_t-\half
\sigma^2 (E_i-E_j)^2 t\right)}. \label{eq:4.27}
\end{eqnarray}
However, the martingale $\{M_{ijt}\}$ defined  for $i\neq j$ by
\begin{eqnarray}
M_{ijt} = \exp\left( \sigma (E_i-E_j)B_t - \half \sigma^2
(E_i-E_j)^2 t\right), \label{eq:4.28}
\end{eqnarray}
which appears in (\ref{eq:4.27}), has the following property:
\begin{eqnarray}
\lim_{t\rightarrow\infty} {\mathbb P}\left( M_{ijt}>0\right) = 0.
\label{eq:4.29}
\end{eqnarray}
In other words, $\{M_{ijt}\}$ converges to zero for large $t$ with
probability one. We note, incidentally, that a geometric Brownian
motion, i.e. a process of the form
\begin{eqnarray}
X_t=\exp\left( \nu B_t - \half \nu^2 t\right),
\end{eqnarray}
has the property that it converges to unity in expectation but to
zero in probability. That is to say,
$\lim_{t\rightarrow\infty}{\mathbb E}[X_t]=1$ whereas
$\lim_{t\rightarrow\infty}{\mathbb P}(X_t>0)=0$. Since
\begin{eqnarray}
H_t = \frac{\pi_j E_j + \sum_{i(\neq j)} \pi_i E_i M_{ijt}}
{\pi_j+\sum_{i(\neq j)} \pi_i M_{ijt}}, \label{eq:4.30}
\end{eqnarray}
we see that $\{H_t\}$ converges to the value $E_j$ with
probability one. A similar argument immediately shows that if
$H=E_j$ then for each value of $i$ we have
\begin{eqnarray}
\lim_{t\rightarrow\infty} \pi_{it}={\bf 1}_{\{i=j\}},
\end{eqnarray}
which shows that $\{|\psi_t\rangle\}$ converges to the L\"uders
state corresponding to the energy eigenvalue $j$ with probability
one, in accordance with the results noted in \cite{brody1}.

The advantage of the expression (\ref{eq:4.14}) is that $\{H_t\}$
and $\{|\psi_t\rangle\}$ are expressed {\sl algebraically} in
terms of the underlying random variable $H$ and the independent
Brownian motion $\{B_t\}$. As a consequence, we can directly
investigate and verify various properties of the reduction process
(\ref{eq:1.1}) without having to resort to numerical integration.

\section{Identification of independent noise and energy}
\label{sec:15}

In solving the stochastic equation (\ref{eq:1.1}) we have
introduced in \S\ref{sec:11} the idea of filtering, that is,
estimation of the value of the random variable $H$, given noisy
information about $H$, where the noise is induced by an
independent random process $\{B_t\}$. Although the method is
useful in obtaining an analytical solution to (\ref{eq:1.1}), the
introduction of these random variables might appear artificial,
because it is not immediately obvious how these variables emerge
out of the problem specified by (\ref{eq:1.1}). Remarkably,
however, it turns out that we can \textit{derive} the quantities
introduced in \S\ref{sec:11} from the ingredients specified in
(\ref{eq:1.1}) and (\ref{eq:1.2}). The aim of this section is to
show how this can be achieved. We start with the following result.

\vspace{0.15cm} \noindent {\bf Proposition 10}. {\it Let $\{H_t\}$
denote the process defined by {\rm (\ref{eq:1.2})} and $\{\xi_t\}$
the process defined by {\rm (\ref{eq:91})}. The random variables
$H_\infty=\lim_{s\to\infty}H_s$ and $B_t=\xi_t-\sigma t H_\infty$
are independent for all $t$. Furthermore, the process $\{B_t\}$ is
a standard Brownian motion.} \vspace{0.15cm}

Proof. We begin by establishing the independence of the random
variables $B_t$ and $H_\infty$. To this end we note that it
suffices to verify that the relation
\begin{eqnarray}
{\mathbb E}\left[\re^{xB_t+yH_\infty}\right]={\mathbb E}\left[
\re^{xB_t}\right]{\mathbb E}\left[\re^{yH_\infty}\right],
\label{eq:11.1}
\end{eqnarray}
holds for all $x,y$. The verification of this property proceeds as
follows. Using the tower property of conditional expectation (see
\S2) we have
\begin{eqnarray}
{\mathbb E}[\re^{xB_t+yH_\infty}] &=& {\mathbb E}\left[{\mathbb E}
\left[ \left. \re^{xB_t+yH_\infty}\right|{\mathcal F}_t^W
\right]\right] \nonumber \\ &=& {\mathbb E}\left[{\mathbb E}
\left[ \left. \re^{x\xi_t+(y-\sigma xt)H_\infty} \right|{\mathcal
F}_t^W\right]\right] \nonumber \\ &=& {\mathbb E}\left[
\re^{x\xi_t}\,{\mathbb E} \left[ \left. \re^{(y- \sigma xt)
H_\infty}\right|{\mathcal F}_t^W\right]\right] , \label{eq:11.2}
\end{eqnarray}
where we have used the $\{{\mathcal F}_t^W\}$-measurability of the
random variable $\xi_t$ in the last step. Let us now consider the
conditional expectation ${\mathbb E} \left[ \left. \re^{(y- \sigma
xt) H_\infty}\right|{\mathcal F}_t^W\right]$ appearing inside the
brackets in (\ref{eq:11.2}). By use of the expression for the
conditional probability of $H_\infty$ obtained in (\ref{eq:4.133})
we deduce that
\begin{eqnarray}
{\mathbb E} \left[ \left. \re^{(y- \sigma xt) H_\infty}\right|
{\mathcal F}_t^W\right] = \frac{\sum_i \pi_i \exp\left((y- \sigma
xt)E_i + \xi_t E_i \sigma -\frac{1}{2}E_i^2\sigma^2t \right)}{
\sum_i \pi_i \exp\left(\xi_t E_i \sigma - \frac{1}{2} E_i^2
\sigma^2t \right)}. \label{eq:11.3}
\end{eqnarray}

To proceed further in determining the outer expectation in
(\ref{eq:11.2}) we make use of the following subsidiary result.

\vspace{0.15cm} \noindent {\bf Lemma 2}. {\it
\begin{eqnarray}
\sum_i \pi_i \exp\left(\xi_t E_i \sigma - \half E_i^2 \sigma^2t
\right) = \exp\left(\sigma\int_0^t H_s \rd \xi_s - \half \sigma^2
\int_0^t H_s^2\rd s\right). \label{eq:11.4}
\end{eqnarray}
} \vspace{0.15cm}

Proof. Let us write $\Phi_t=\Phi(\xi_t,t)$ for the left-hand side
of (\ref{eq:11.4}), where $\Phi(\xi,t)$ is the function of two
variables defined by
\begin{eqnarray}
\Phi(\xi,t)=\sum_i \pi_i \exp\left(\xi E_i \sigma - \half E_i^2
\sigma^2t \right) .
\end{eqnarray}
Then by Ito's lemma we have
\begin{eqnarray}
\rd\Phi_t = {\dot\Phi}_t\rd t+\Phi^\prime_t \rd \xi_t + \half
\Phi^{\prime\prime}_t (\rd\xi_t)^2,
\end{eqnarray}
where the dot and the prime denote differentiation with respect to
$t$ and $\xi$, respectively. Next we observe that $(\rd\xi_t)^2
=\rd t$, that ${\dot\Phi}_t+\frac{1}{2}\Phi^{\prime\prime}_t=0$,
and that $\Phi^{\prime}_t=\sigma H_t\Phi_t$, the last relation
following from (\ref{eq:4.14}). As a consequence we see that
$\{\Phi_t\}$ satisfies
\begin{eqnarray}
\rd\Phi_t = \sigma H_t \Phi_t \rd \xi_t.
\end{eqnarray}
Finally we note that the integral representation for this
stochastic differential equation, with initial condition
$\Phi_0=1$, is given by the right-hand side of (\ref{eq:11.4}).
\hspace*{\fill} $\diamondsuit$

The key point is that the right-hand side of (\ref{eq:11.4}) can
be used as change-of-measure density. Recall that $\{W_t\}$ is a
standard Brownian motion in the measure ${\mathbb P}$. Since
$\{H_t\}$ is bounded and $\{{\mathcal F}_t^W\}$-adapted, it
follows by Girsanov's theorem that there exists an equivalent
probability measure ${\mathbb Q}$ such that the process
$\{\xi_t\}$ defined by
\begin{eqnarray}
\xi_t = W_t + \sigma\int_0^t H_s \rd s \label{eq:11.45}
\end{eqnarray}
is a standard Brownian motion in the ${\mathbb Q}$-measure. We let
$\Phi_t$ denote the change-of-measure density in the right-hand
side of (\ref{eq:11.4}). Then for any $\{{\mathcal
F}_t^W\}$-measurable random variable $X_t$ the conditional
expectations in these two probability measures are related
according to the scheme
\begin{eqnarray}
{\mathbb E}_s^{\mathbb P}[X_t]=\frac{1}{\Phi_s}{\mathbb
E}_s^{\mathbb Q} [\Phi_t X_t], \quad {\rm and}\quad {\mathbb
E}_s^{\mathbb Q}[X_t] = \Phi_s{\mathbb E}_s^{\mathbb P}
[\frac{1}{\Phi_t} X_t]. \label{eq:11.5}
\end{eqnarray}

Equipped with these results we proceed to determine the
conditional expectation (\ref{eq:11.2}). In particular, if we
substitute (\ref{eq:11.3}) in (\ref{eq:11.2}) and use the fact
that the denominator appearing in the expectation is the
change-of-measure density $\Phi_t$, and hence $\{\xi_t\}$ is a
standard Brownian motion in the ${\mathbb Q}$-measure, we can
apply the second identity in (\ref{eq:11.5}) to deduce that
\begin{eqnarray}
{\mathbb E}[\re^{xB_t+yH_\infty}] &=& {\mathbb E}^{\mathbb
Q}\left[ \re^{x\xi_t} \sum_i \pi_i\,\re^{(y- \sigma xt)E_i + \xi_t
E_i \sigma -\frac{1}{2} E_i^2\sigma^2t}\right] \nonumber \\ &=&
\sum_i \pi_i\,\re^{(y- \sigma xt)E_i-\frac{1}{2}
E_i^2\sigma^2t}\,{\mathbb E}^{\mathbb Q}\left[ \re^{(x+E_i\sigma)
\xi_t}\right] \nonumber \\ &=& \sum_i \pi_i\,\re^{(y- \sigma xt)
E_i-\frac{1}{2} E_i^2\sigma^2t}\,\re^{\frac{1}{2}(x+E_i\sigma)^2t}
\nonumber \\ &=& \Big(\sum_i \pi_i\, \re^{yE_i}\Big)
\re^{\frac{1}{2}x^2 t} . \label{eq:11.6}
\end{eqnarray}
This establishes the relation (\ref{eq:11.1}), and hence that
random variables $B_t$ and $H_\infty$ are independent. In
addition, as a bonus the result
\begin{eqnarray}
{\mathbb E}[\re^{xB_t}] = \re^{\frac{1}{2}x^2 t} \label{eq:11.7}
\end{eqnarray}
shows that $B_t$ is normally distributed with mean zero and
variance $t$.

To complete the proof that $\{B_t\}$ is a standard Brownian motion
we are required, in addition to establishing its normality, to
verify that the process $\{B_t\}$ has independent increments.
Alternatively, it suffices to demonstrate that
\begin{eqnarray}
{\mathbb E}\left[\re^{xB_t+y(B_T-B_t)}\right]={\mathbb E}\left[
\re^{xB_t}\right]{\mathbb E}\left[\re^{y(B_T-B_t)}\right]
\label{eq:11.8}
\end{eqnarray}
for any nonzero constants $x,y$. Using the definition for
$\{B_t\}$ and the tower property of conditional expectation we can
write
\begin{eqnarray}
{\mathbb E}\left[\re^{xB_t+y(B_T-B_t)}\right] &=& {\mathbb E}
\left[ \re^{(x-y)\xi_t+y\xi_T}\, \re^{-\left(x\sigma t+y\sigma
(T-t)\right)H_\infty}\right] \nonumber \\ &=& {\mathbb E}\left[
\re^{(x-y)\xi_t+y\xi_T}\,{\mathbb E}\left[\left.\re^{-\left(x
\sigma t+y\sigma (T-t)\right)H_\infty}\right|{\mathcal F}_T^{W}
\right] \right]. \label{eq:11.9}
\end{eqnarray}
Once again from (\ref{eq:4.133}) we have
\begin{eqnarray}
{\mathbb E}\left[\left.\re^{-\left(x \sigma t+y\sigma
(T-t)\right)H_\infty}\right|{\mathcal F}_T^{W} \right] =
\frac{\sum_i\pi_i\,\re^{-x\sigma tE_i-y\sigma(T-t)E_i+\xi_T E_i
\sigma -\frac{1}{2}E_i^2\sigma^2T}}{ \sum_i \pi_i \re^{\xi_T E_i
\sigma - \frac{1}{2} E_i^2 \sigma^2 T}} \label{eq:11.10}
\end{eqnarray}
for the inner expectation in (\ref{eq:11.9}). Substituting
(\ref{eq:11.10}) in (\ref{eq:11.9}) and noting the fact that the
denominator in the expectation is the change-of-measure density
$\Phi_T$ we deduce, after some rearrangement of terms, that
\begin{eqnarray}
\fl{\mathbb E}\left[\re^{xB_t+y(B_T-B_t)}\right] &=&\sum_i \pi_i\,
\re^{-x\sigma tE_i-y\sigma(T-t) E_i-\frac{1}{2} E_i^2 \sigma^2T}\,
{\mathbb E}^{\mathbb Q}\left[ \re^{(x-y) \xi_t +
(y+E_i\sigma)\xi_T} \right] \nonumber \\ &=& \re^{\frac{1}{2}x^2t
+ \frac{1}{2}y^2(T-t)}. \label{eq:11.11}
\end{eqnarray}
Here we have made use of the Gaussian property
\begin{eqnarray}
{\mathbb E}^{\mathbb Q}\left[ \re^{a\xi_t+b\xi_T}\right]&=&\exp
\left( \half {\mathbb E}^{\mathbb Q}[(a\xi_t+b\xi_T)^2]\right)
\nonumber \\ &=& \exp\left(\half(a^2+2ab)t+\half b^2 T\right)
\end{eqnarray}
satisfied by the random variables $\xi_t$ and $\xi_T$ in the
${\mathbb Q}$-measure. The result of (\ref{eq:11.11}) establishes
(\ref{eq:11.8}), and thus we conclude that the process $\{B_t\}$
is normally distributed with zero mean and variance $t$, and has
independent increments. Therefore, $\{B_t\}$ is a standard
Brownian motion. \hspace*{\fill} $\diamondsuit$

\section{Finite-time collapse model}
\label{sec:16}

In the foregoing sections we have investigated the properties of
energy-based collapse models for which state reduction is achieved
{\it asymptotically} in time. That is to say, although for a
suitable choice of the parameter $\sigma$ the state reaches the
close vicinity of one of the energy eigenstates virtually
instantaneously, for a strict collapse for which the variance
process $\{V_t\}$ vanishes identically, we must take the limit
$t\to\infty$. There are circumstances, however, in which it might
be preferable to formulate a model that achieves strict collapse
in finite time duration. An example for such a model has been
proposed recently~\cite{brody6}. In what follows we shall apply
the methodologies developed above to work out the properties of
finite-time collapse models.

The model that we consider here, which gives rise to a finite-time
collapse, is given by the following stochastic equation:
\begin{eqnarray}
\fl\rd |\psi_t\rangle = -\ri{\hat H}|\psi_t\rangle\rd t - \octa
\left(\frac{\sigma T}{T-t}\right)^2 ({\hat H}-H_t )^2|\psi_t
\rangle\rd t +\half \frac{\sigma T}{T-t} ({\hat
H}-H_t)|\psi_t\rangle\rd W_t . \label{eq:16.1}
\end{eqnarray}
We deduce immediately from the discussion in \S\ref{sec:7} that
the dynamical law (\ref{eq:16.1}) preserves the norm
$\langle\psi_t|\psi_t \rangle$ of the state, and that the
associated energy process $\{H_t\}$ is a martingale. In
particular, a short calculation making use of the Ito calculus
shows that the energy process satisfies
\begin{eqnarray}
\rd H_t = \sigma_t V_t\,\rd W_t, \label{eq:196}
\end{eqnarray}
where we have defined, for convenience, the deterministic function
$\{\sigma_t\}$ by
\begin{eqnarray}
\sigma_t = \frac{\sigma T}{T-t},
\end{eqnarray}
and $\{V_t\}$ is the associated variance process. Note that
(\ref{eq:16.1}) can be obtained from (\ref{eq:1.1}) by the
substitution $\sigma\to \sigma_t$. Thus, (\ref{eq:16.1}) contains
two freely specifiable parameters, namely, $\sigma$ and $T$. The
latter will be identified with the time at which the collapse is
completed. More generally, we may regard the collapse time $T$ as
a random variable having some density $p(T)$ defined on the
positive real line. Then the collapse time $T$ itself becomes
random; the analysis of this case will be pursued elsewhere. Here
we shall treat $T$ as a fixed parameter.

In order to solve (\ref{eq:16.1}) we shall make an ansatz
analogous to the one introduced in (\ref{eq:4.1}). Now from the
point of view of filtering theory, the collapse of the state in
the model (\ref{eq:1.1}) takes place only asymptotically because
the `noise to signal' ratio, whose magnitude is of order
$\sqrt{t}/t$, vanishes only asymptotically as $t\to\infty$.
Therefore, in order to achieve a finite time collapse we consider
the use of a \emph{Brownian bridge} as the source for the noise. A
Brownian bridge with duration $T$ can be regarded as a standard
Brownian motion constrained to take value zero at time $t=0$ and
also at time $t=T$. By using a Brownian bridge as the source for
the noise, the value of the unknown random variable $H$ will be
revealed in finite time $T$, since the contribution of noise
vanishes at that time. Specifically, the magnitude of noise to
signal ratio is given by $\sqrt{(T-t)/tT}$, which vanishes as
$t\to T$. It remains to be shown that the solution of such a
filtering problem corresponds to the solution of a finite time
collapse model (\ref{eq:16.1}). In what follows we shall
demonstrate that this is the case.

We thus consider the information process $\{\xi_t\}_{0\leq t\leq
T}$ defined in this case by
\begin{eqnarray}
\xi_t = \sigma t H + B_t - \frac{t}{T} B_T, \label{eq:198}
\end{eqnarray}
where $\sigma$ is a constant, $\{B_t\}$ is a standard Brownian
motion, and $H$ is a discrete random variable taking the values
$\{E_i\}$ with probability $\{\pi_i\}$. It is evident from
definition that $\xi_0=0$ and that $\xi_t/\sigma t=H$. The process
$\{\beta_t\}_{0\leq t\leq T}$ defined by the combination
\begin{eqnarray}
\beta_t = B_t - \frac{t}{T} B_T \label{eq:16.3-5}
\end{eqnarray}
is a standard Brownian bridge on the interval $t\in[0,T]$
satisfying $\beta_0=0$ and $\beta_T=0$. We assume that $H$ and
$\{\beta_t\}$ are independent. It should be evident from the
definition (\ref{eq:16.3-5}) that a Brownian bridge is normally
distributed with mean ${\mathbb E}[\beta_t]=0$ and covariance
\begin{eqnarray}
{\rm Cov}[\beta_s,\beta_t] &=& {\mathbb E}\left[ B_s B_t -
\frac{1}{T}(sB_t+tB_s)B_T+\frac{1}{T^2}stB_T^2\right] \nonumber \\
&=& s\left( 1-\frac{t}{T}\right) \label{eq:199}
\end{eqnarray}
for $s\leq t$. In deriving (\ref{eq:199}) we have made use of the
independent increments property satisfied by $\{B_t\}$ to deduce
that ${\mathbb E}[B_sB_t]={\mathbb E}[B_s(B_t-B_s+B_s)]=s$. The
Brownian bridge, on the other hand, does not possess independent
increments.

Our objective now, as before, is to determine the best estimate
for the variable $H$ given the information concerning the
trajectory $\{\xi_u\}_{ 0\leq u\leq t}$ of the process $\{\xi_u\}$
from time $u=0$ to time $u=t\leq T$. In particular, the conclusion
of Proposition~7 remains valid in the present context: that is to
say, the estimate that minimises the quadratic error is given by
the conditional expectation ${\mathbb E}[H|{\mathcal F}_t^\xi]$.
To calculate this conditional expectation we shall make use of the
following key result:

\vspace{0.15cm} \noindent {\bf Lemma 3}. {\it Let $\xi_t = \sigma
t H + \beta_t$, where $H$ is a random variable taking the values
$E_i$ $(i=1,2,\ldots,N)$ with probability ${\mathbb P} (H=E_i)
=\pi_i$, $\sigma$ is a constant, and $\{\beta_t\}$ is a standard
${\mathbb P}$-Brownian bridge on the interval $t\in[0,T]$,
independent of $H$. Then $\{\xi_t\}_{0\leq t\leq T}$ is a Markov
process. } \vspace{0.15cm}

Proof of Lemma 3. To show that $\{\xi_t\}$ is Markovian, we must
show that
\begin{eqnarray}
{\mathbb P}( \xi_t\leq x|{\mathcal F}_s^\xi)= {\mathbb P}( \xi_t
\leq x|\xi_s)
\end{eqnarray}
for all $x\in{\mathbb R}$ and all $s,t$ such that $0\leq s\leq
t\leq T$. It will suffice to verify that
\begin{eqnarray}
{\mathbb P}\left( \xi_t\leq x|\xi_s, \xi_{s_1}, \xi_{s_2}, \cdots,
\xi_{s_k}\right) ={\mathbb P}\left( \xi_t\leq x|\xi_s\right)
\end{eqnarray}
for any times $t,s,s_1,s_2,\ldots,s_k$ such that $T\geq
t>s>s_1>s_2>\cdots>s_k>0$. We remark that for any times $t,s,s_1$
satisfying $t>s>s_1$ the random variables $\beta_{t}$ and
$\beta_{s}/s-\beta_{s_1}/s_1$ have vanishing covariance, and thus
are independent. More generally, for $s>s_1>s_2>s_3$ the random
variables $\beta_{s}/s-\beta_{s_1} /s_1$ and $\beta_{s_2}/s_2-
\beta_{s_3}/s_3$ are independent. We note that $\xi_s/s-
\xi_{s_1}/s_1=\beta_{s}/s - \beta_{s_1} /s_1$. It follows that
\begin{eqnarray}
\fl{\mathbb P}\left( \xi_t \leq x|\xi_s, \xi_{s_1}, \xi_{s_2},
\cdots, \xi_{s_k}\right) &=&{\mathbb P}\left( \xi_t\leq
x\Big|\xi_s, \frac{\xi_s}{s}-\frac{\xi_{s_1}} {s_1},
\frac{\xi_{s_1}}{s_1}- \frac{\xi_{s_2}}{s_2}, \cdots,
\frac{\xi_{s_{k-1}}}{s_{k-1}}- \frac{\xi_{s_k}}{s_k}\right)
\nonumber \\ && \hspace{-4.0cm} = {\mathbb P}\left( \xi_t\leq
x\Big|\xi_s, \frac{\beta_{s}}{s}-\frac{\beta_{s_1}}{s_1},
\frac{\beta_{s_1}}{s_1}- \frac{\beta_{s_2}}{s_2}, \cdots,
\frac{\beta_{s_{k-1}}}{s_{k-1}}- \frac{\beta_{s_k}}{s_k}\right).
\end{eqnarray}
Since $\xi_s$ and $\xi_t$ are independent of $\beta_{s}/s -
\beta_{s_1} /s_1$, $\beta_{s_1}/s_1 - \beta_{s_2} /s_2$, $\cdots$,
$\beta_{s_{k-1}}/s_{k-1}- \beta_{s_k}/s_k$, the desired result
follows immediately. \hspace*{\fill} $\diamondsuit$

Because $\{\xi_t\}$ is a Markov process, the conditional
expectation ${\mathbb E}[H|{\mathcal F}_t^\xi]$ simplifies to
$H_t={\mathbb E}\left[H|\xi_t\right]$ so that we only need to
specify the value $\xi_t$ of the process at time $t$, and not the
entire trajectory $\{\xi_u\}_{ 0\leq u\leq t}$. We shall first
establish the following result.

\vspace{0.15cm} \noindent {\bf Proposition 11}. {\it Let $H$ be a
random variable taking the value $E_i$ with probability $\pi_i$
$(i=1,2,\ldots,n)$, and set $\xi_t= \sigma t H+B_t-(t/T)B_T$ for
$0\leq t<T$, where $\sigma$ is a constant and the Brownian motion
$\{B_t\}$ is independent of $H$. Then the conditional expectation
$H_t = {\mathbb E}[ H|{\mathcal F}_t^\xi]$ is given by
\begin{eqnarray}
H_t = \frac{\sum_i \pi_i E_i \exp\left(\frac{\sigma\xi_t E_iT-
\frac{1}{2}\sigma^2 E_i^2 t T}{T-t}\right)}{\sum_i \pi_i
\exp\left(\frac{\sigma \xi_t E_iT- \frac{1}{2}\sigma^2 E_i^2 t
T}{T-t}\right)} . \label{eq:16.4}
\end{eqnarray}
} \vspace{0.15cm}

Proof. The conditional expectation $H_t={\mathbb E}\left[
H|\xi_t\right]$ can be expressed in terms of the conditional
probability as follows:
\begin{eqnarray}
H_t = \sum_i E_i\,{\mathbb P}(H=E_i|\xi_t). \label{eq:16.5}
\end{eqnarray}
To determine the conditional probability ${\mathbb P} (H=E_i|\xi)$
we note that according to the Bayes formula that we can write
\begin{eqnarray}
{\mathbb P}(H=E_i|\xi_t) = \frac{\pi_i \rho(\xi_t|H=E_i)} {\sum_i
\pi_i \rho(\xi_t|H=E_i)}, \label{eq:16.6}
\end{eqnarray}
where
\begin{eqnarray}
\rho(\xi_t|H=E_i) = \sqrt{{\textstyle\frac{T}{2\pi t(T-t)}}}
\exp\left( -\frac{(\xi_t-\sigma t E_i)^2T}{2t(T-t)}\right).
\label{eq:16.7}
\end{eqnarray}
Expression (\ref{eq:16.7}) follows from the fact that conditional
on $H=E_i$ the variable $\xi_t$ in (\ref{eq:198}) is normally
distributed with mean $\sigma t E_i$ and variance
\begin{eqnarray}
{\mathbb E}\left[ (B_t-(t/T)B_T)^2\right]=t(T-t)/T.
\label{eq:16.8}
\end{eqnarray}
Putting these together, we deduce (\ref{eq:16.4}) after some
rearrangements of terms. \hspace*{\fill} $\diamondsuit$

From the expression (\ref{eq:16.4}) we can infer directly the
property that $H_t\to E_k$ as $t\to T$, provided we set $H=E_k$.
Writing $H_t^k$ for the conditional energy process $H_t(H=E_k)$,
$\omega_{ij}=E_i-E_j$ for the difference of energy eigenvalues,
and substituting $\xi_t=\xi_t^k=\sigma t E_k+\beta_t$ in
(\ref{eq:16.4}), we obtain
\begin{eqnarray}
H_{t}^k &=& \frac{\sum_i \pi_i E_i \exp\left(\frac{\sigma\xi_t^k
E_iT- \frac{1}{2}\sigma^2 E_i^2 t T}{T-t}\right)}{\sum_i \pi_i
\exp \left(\frac{\sigma \xi_t^k E_iT- \frac{1}{2}\sigma^2 E_i^2 t
T} {T-t}\right)}\nonumber \\ &=& \frac{\pi_kE_k+\sum_{i\neq
k}\exp\left(\frac{\sigma T\omega_{ik} \beta_t-\frac{1}{2} \sigma^2
\omega_{ik}^2tT}{T-t}\right)} {\pi_k +\sum_{i\neq k}\exp\left(
\frac{\sigma T\omega_{ik} \beta_t - \frac{1}{2} \sigma^2
\omega_{ik}^2tT}{T-t}\right)}. \label{eq:16.9}
\end{eqnarray}
Observe that for each $i$ the numerator in the exponent in
(\ref{eq:16.9}) approaches a strictly negative number $-
\frac{1}{2}\sigma^2\omega_{ik}^2tT$. Hence as $t\to T$ all the
exponential terms are rapidly suppressed and we are left with the
desired outcome: $H_T^k = E_k$.

\section{Innovation process for finite-time collapse model}
\label{sec:17}

Let us analyse the properties of the process $\{H_t\}$ in
(\ref{eq:16.4}) more closely. By taking the stochastic
differential of (\ref{eq:16.4}) we obtain
\begin{eqnarray}
\rd H_t = \sigma_t V_t \left[ \frac{1}{T-t} \Big( \xi_t-\sigma T
H_t\Big) \rd t + \rd\xi_t \right] , \label{eq:17.1}
\end{eqnarray}
where $V_t={\mathbb E}[H^2|{\mathcal F}_t^\xi]-H_t^2$ is the
conditional variance of the random variable $H$. Clearly, there
exists a choice of a process $\{W_t\}$ defined in terms of
$\{H_t\}$ and $\{\xi_t\}$ such that the drift term in the
dynamical equation (\ref{eq:17.1}) can be removed, when expressed
in terms of $\{W_t\}$. It remains to be shown that such a process
is a Brownian motion that derives the dynamics of the state
(\ref{eq:16.1}). We shall proceed by verifying the following
result.

\vspace{0.15cm} \noindent {\bf Proposition 12}. {\it The process
$\{W_t\}$ defined by
\begin{eqnarray}
W_t = \int_0^t \frac{1}{T-s}\Big(\xi_s-\sigma TH_s\Big)\rd s +
\xi_t \label{eq:17.2}
\end{eqnarray}
is an $\{{\mathcal F}_t^\xi\}$-Brownian motion.} \vspace{0.15cm}

Proof. First we note that the tower property of conditional
expectation shows
\begin{eqnarray}
{\mathbb E}[B_t|{\mathcal F}_s^\xi] = {\mathbb E}\left[ \left.
{\mathbb E}[ B_t|{\mathcal F}_t^B,H]\right|{\mathcal F}_s^\xi
\right] = {\mathbb E}_s\left[B_s\right], \label{eq:xxx11}
\end{eqnarray}
where we write ${\mathbb E}_s[-]={\mathbb E}[-|{\mathcal
F}_s^\xi]$. It follows from (\ref{eq:198}) that $\xi_t={\mathbb
E}_t[\xi_t]$, and hence that $\xi_t$ is given by
\begin{eqnarray}
\xi_t=\sigma t H_t+\left(1-\frac{t}{T}\right){\mathbb E}_t[B_t].
\label{eq:xxxxx}
\end{eqnarray}
We now proceed to establish that $\{W_t\}$ as defined by
(\ref{eq:17.2}) is an $\{{\mathcal F}_t^\xi\}$-martingale. For
$t\leq u$ we have:
\begin{eqnarray}
\fl \hspace{1.4cm}{\mathbb E}_t[W_u] &=& {\mathbb E}_t[\xi_u] +
{\mathbb E}_t\left[ \int_0^u \frac{1}{T-s}\Big(\xi_s-\sigma T
H_s\Big)\rd s \right] \nonumber \\ &=& {\mathbb E}_t[\xi_u] +
\int_0^t \frac{1}{T-s}(\xi_s-\sigma T H_s)\rd s + \int_t^u
\frac{1}{T-s}({\mathbb E}_t[\xi_s]-\sigma T H_t)\rd s.
\end{eqnarray}
Here we have used the fact that $\{\xi_t\}$ and $\{H_t\}$ are
$\{{\mathcal F}_t^\xi\}$-adapted, and that ${\mathbb E}_t[H_s]=
H_t$ for $t\leq s$. Therefore,
\begin{eqnarray}
\fl \hspace{2cm}{\mathbb E}_t[W_u] &=& {\mathbb E}_t\left[\sigma u
H + B_u - \frac{u}{T}B_T\right] + W_t-\xi_t \nonumber \\ && +
\int_t^u \frac{1}{T-s}{\mathbb E}_t\left[\sigma s H + B_s -
\frac{s}{T}B_T\right]\rd s -\sigma T H_t \int_t^u \frac{1}{T-s}\rd
s \nonumber \\ &=& \sigma u H_t + W_t-\xi_t +\sigma H_t \int_t^u
\frac{s}{T-s}\,\rd s -\sigma T H_t \int_t^u \frac{1}{T-s}\,\rd s
\nonumber \\ && + {\mathbb E}_t [B_t] \left( 1-\frac{u}{T} +
\int_t^u \frac{1}{T-s}\left( 1- \frac{s}{T} \right)\,\rd s\right)
\nonumber \\ &=& W_t-\xi_t+ \sigma t H_t + {\mathbb E}_t
[B_t]\left(1-\frac{t}{T}\right) \nonumber \\ &=& W_t,
\end{eqnarray}
where in the final step we have made use of the relation
(\ref{eq:xxxxx}). This establishes the martingale property
satisfied by $\{W_t\}$. On the other hand, (\ref{eq:17.2}) implies
$(\rd W_t)^2=(\rd\xi_t)^2$, whereas (\ref{eq:198}) implies
$(\rd\xi_t)^2=\rd t$. It follows that $(\rd W_t)^2=\rd t$, and
this establishes the assertion that $\{W_t\}$ is an $\{{\mathcal
F}_t^\xi\}$-Brownian motion.  \hspace*{\fill} $\diamondsuit$

We remark that by substituting (\ref{eq:17.2}) in (\ref{eq:17.1})
we obtain the dynamics
\begin{eqnarray}
\rd H_t = \sigma_t V_t\,\rd W_t \label{eq:17.3}
\end{eqnarray}
for the process $\{H_t\}$ given in (\ref{eq:16.4}), which shows
that $\{H_t\}$ is a martingale. On the other hand, by taking the
stochastic differential of the energy process
$H_t=\langle\psi_t|{\hat H}|\psi_t\rangle/\langle\psi_t|\psi_t
\rangle$ using (\ref{eq:16.1}) we have obtained (\ref{eq:196}). To
show that (\ref{eq:16.4}) is indeed the energy process associated
with the dynamics (\ref{eq:16.1}) we must demonstrate that the two
processes labelled by $\{W_t\}$ are identical. In particular, we
have the following result.

\vspace{0.15cm} \noindent {\bf Proposition 13}. {\it The
innovation process $\{W_t\}$ defined in {\rm (\ref{eq:17.2})} is
the Brownian motion that derives the dynamics of the wave function
in {\rm (\ref{eq:16.1})}.} \vspace{0.15cm}

Proof. The stochastic differential equation (\ref{eq:16.1}) can be
given the following integral representation
\begin{eqnarray}
\fl|\psi_t\rangle = \exp\left( -{\ri}{\hat H}t - \quat \int_0^t
\sigma_s^2({\hat H}-H_s)^2{\rd}s + \half \int_0^t \sigma_s ({\hat
H}-H_s) {\rd}W_s \right) |\psi_0\rangle. \label{eq:17.4}
\end{eqnarray}
This can be expressed more concisely as $|\psi_t\rangle = {\hat
U}_t {\hat R}_t |\psi_0\rangle$, where
\begin{eqnarray}
{\hat U}_t = \exp\Big( -{\ri}{\hat H}t\Big) \label{eq:17.5}
\end{eqnarray}
is the usual unitary evolution operator, and
\begin{eqnarray}
{\hat R}_t = \exp\left( \half \int_0^t\sigma_s ({\hat H}-H_s)
{\rd}W_s - \quat \int_0^t\sigma_s^2 ({\hat H}-H_s)^2{\rd}s \right)
\label{eq:17.6}
\end{eqnarray}
is the `reduction' operator. The square of ${\hat R}_t$, which we
denote by ${\hat M}_t$, is an operator-valued martingale, given by
\begin{eqnarray}
{\hat M}_t &=& \exp\left( \int_0^t\sigma_s({\hat H}-H_s) {\rd}W_s
- \half \int_0^t\sigma^2_s ({\hat H} -H_s)^2 {\rd}s \right)
\nonumber \\ &=& \frac{\exp\left(\int_0^t{\hat H}\sigma_s({\rd}W_s
+ \sigma_s H_s{\rd}s)-\frac{1}{2} \int_0^t \sigma_s^2{\hat H}^2
{\rd}s \right)}{\exp\left( \int_0^t H_s\sigma_s ({\rd}W_s+\sigma_s
H_s{\rd}s)-\frac{1}{2} \int_0^t \sigma_s^2 H_s^2 {\rd}s \right)}.
\label{eq:17.7}
\end{eqnarray}
Let us now introduce a modified Brownian motion $\{W_t^*\}$
according to
\begin{eqnarray}
W_t^* = W_t + \int_0^t \sigma_s H_s {\rd}s, \label{eq:17.8}
\end{eqnarray}
so $\rd W_t^*={\rd}W_t+\sigma_t H_t{\rd}t$. While $\{W_t^*\}$ is a
drifted Brownian motion in the probability measure ${\mathbb P}$,
we can construct another probability measure ${\mathbb Q}$ in
which the process $\{W_t^*\}$ becomes a standard Brownian motion.
Then, because ${\hat H}$ is constant in time, we can write ${\hat
M}_t$ in the simple form
\begin{eqnarray}
{\hat M}_t = \frac{1}{\Phi_t} \exp\left( {\hat H}\int_0^t\sigma_s
\rd W_s^* - \half {\hat H}^2\int_0^t\sigma_s^2 \rd s \right),
\label{eq:17.9}
\end{eqnarray}
where
\begin{eqnarray}
\Phi_t = \exp\left( \int_0^t \sigma_s H_s {\rd}W_s^* - \half
\int_0^t \sigma_s^2 H_s^2 {\rd}s \right) \label{eq:17.10}
\end{eqnarray}
is a positive martingale process.

Recall that (\ref{eq:16.1}) preserves the norm of $|\psi_0
\rangle$. Therefore, if we assume initially that $\langle
\psi_0|\psi_0 \rangle=1$, then it follows that
$\langle\psi_0|{\hat M}_t |\psi_0\rangle=1$ for all $t$. Thus we
deduce from (\ref{eq:17.7}) and (\ref{eq:17.8}) that
\begin{eqnarray}
\Phi_t = \langle\psi_0| \exp\left( {\hat H}\!\int_0^t\!\sigma_s
\rd W_t^* - \half {\hat H}^2\!\int_0^t\!\sigma_s^2 \rd t \right)
|\psi_0\rangle . \label{eq:10.17}
\end{eqnarray}
As a consequence we can write
\begin{eqnarray}
{\hat M}_t = \frac{\exp\left({\hat H}\int_0^t\sigma_s\rd W_t^* -
\frac{1}{2} {\hat H}^2\int_0^t\sigma_s^2\rd t\right)} {\langle
\psi_0| \exp\left( {\hat H} \int_0^t  \sigma_s \rd W_t^* -
\frac{1}{2} {\hat H}^2 \int_0^t \sigma_s^2 \rd t \right)
|\psi_0\rangle} , \label{eq:10.18}
\end{eqnarray}
which has the effect of isolating the dependence of ${\hat M}_t$
on $\{H_t\}$. In particular, ${\hat M}_t$ depends on $\{H_t\}$
entirely through the modified Brownian motion $\{W_t^*\}$. The
process $\{H_t\}$ in turn is given by $H_t=\langle\psi_t|{\hat
H}|\psi_t\rangle/\langle\psi_t|\psi_t \rangle$, from which it
follows that $H_t = \langle\psi_0|{\hat H}{\hat M}_t|\psi_0
\rangle$. Therefore, by use of (\ref{eq:10.18}) we have
\begin{eqnarray}
H_t = \frac{\langle{\psi}_0| {\hat H} \exp\left({\hat H}
\int_0^t\!\sigma_s\rd W_s^*\!-\!\frac{1}{2} {\hat H}^2\int_0^t\!
\sigma_s^2 \rd s\right)|\psi_0\rangle} {\langle\psi_0| \exp\left(
{\hat H}\int_0^t\!\sigma_s\rd W_s^*\!-\!\frac{1}{2} {\hat H}^2
\int_0^t\!\sigma_s^2 \rd s \right)|\psi_0\rangle}
,\label{eq:10.19}
\end{eqnarray}
which shows that $H_t$ can be expressed in terms of $\{W_t^*\}$
and $t$. This is given by
\begin{eqnarray}
H_t = \frac{\sum_i \pi_i E_i \exp\left(E_i\int_0^t\sigma_s\rd
W_s^* - \frac{1}{2} E_i^2 \int_0^t\sigma_s^2 \rd s\right)}{\sum_i
\pi_i \exp\left( E_i \int_0^t\sigma_s\rd W_s^* -\frac{1}{2} E_i^2
\int_0^t \sigma_s^2 \rd s\right)} , \label{eq:10.20}
\end{eqnarray}
where $\pi_i$ denotes the initial probability that the eigenvalue
attained is $E_i$.

Now if the process $\{H_t\}$ obtained in (\ref{eq:16.4}) is the
energy process (\ref{eq:10.20}), then from the relation $\int_0^t
\sigma_s^2\rd s=\sigma^2tT/(T-t)$, we deduce, by comparison of
(\ref{eq:16.4}) and (\ref{eq:10.20}), that
\begin{eqnarray}
\xi_t = (T-t)\int_0^t \frac{1}{T-s}\,\rd W_s^*  \label{eq:10.21}
\end{eqnarray}
must be satisfied. To show that (\ref{eq:10.21}) is satisfied, we
remark first that the stochastic differential of (\ref{eq:17.2})
is given by
\begin{eqnarray}
\rd W_t + \sigma T \frac{1}{T-t}H_t\rd t = \frac{1}{T-t}\xi_t\rd t
+ \rd \xi_t . \label{eq:10.22}
\end{eqnarray}
On the other hand, the differential form of (\ref{eq:17.8}) is
\begin{eqnarray}
\rd W_t + \sigma T \frac{1}{T-t}H_t\rd t = \rd W_t^* .
\label{eq:10.23}
\end{eqnarray}
Therefore, by comparing (\ref{eq:10.22}) and (\ref{eq:10.23}) we
deduce the relation
\begin{eqnarray}
\rd \xi_t = -\frac{1}{T-t}\xi_t\rd t + \rd W_t^* .
\end{eqnarray}
This, however, is the differential form of (\ref{eq:10.21}). It
follows that the process $\{H_t\}$ obtained in (\ref{eq:16.4}) is
the energy process (\ref{eq:10.20}) associated with the collapse
model (\ref{eq:16.1}). In particular, the process $\{W_t\}$
defined in (\ref{eq:17.2}) is the Brownian motion that drives the
dynamics of the state (\ref{eq:16.1}). \hspace*{\fill}
$\diamondsuit$

The above result also shows that the process $\{\xi_t\}$ defined
by (\ref{eq:198}) is itself a Brownian bridge in the ${\mathbb
Q}$-measure. This follows from the integral representation
(\ref{eq:10.21}) above, which shows that in the ${\mathbb
Q}$-measure, under which $\{W_t^*\}$ is a standard Brownian
motion, $\{\xi_t\}$ is a zero-mean Gaussian process with
autocovariance given by
\begin{eqnarray}
\fl \hspace{1.3cm} {\mathbb E}\left[\xi_s\xi_t\right] &=& (T-s)
(T-t) {\mathbb E} \left[ \int_0^s \frac{1}{T-u}\,\rd W_u^*
\int_0^t \frac{1}{T-v} \,\rd W_v^* \right] \nonumber \\ &=&
{\mathbb E} \left[ \left( \int_0^s \frac{1}{T-u}\,\rd
W_u^*\right)^2\right] + {\mathbb E} \left[ \int_0^s
\frac{1}{T-u}\,\rd W_u^* \int_s^t \frac{1}{T-u} \,\rd W_u^*\right]
\nonumber \\ &=& {\mathbb E} \left[ \left( \int_0^s
\frac{1}{T-u}\,\rd W_u^*\right)^2\right] + {\mathbb E} \left[
\int_0^s \frac{1}{T-u}\,\rd W_u^*\right] {\mathbb E} \left[
\int_s^t \frac{1}{T-u}\,\rd W_u^*\right] \nonumber \\ &=& s\left(
1-\frac{t}{T} \right) \label{eq:cov3}
\end{eqnarray}
for $s\leq t$. Here we have substituted the integral
representation (\ref{eq:10.21}) in the right-side of
(\ref{eq:cov3}), applied the Wiener-Ito isometry and used the
independent increments property of Brownian motion. We shall make
use of this result to establish Proposition~14 below.

\section{Reverse-construction for finite-time collapse model}
\label{sec:18}

We have demonstrated in the previous section that the closed-form
solution to the stochastic equation (\ref{eq:16.1}) can be
obtained by use of a nonlinear filtering methodology, in which we
have introduced a pair of independent random data $H$ and
$\{\beta_t\}$. Conversely, starting from the stochastic equation
(\ref{eq:16.1}), we can derive the existence of such a pair of
independent data. We let $\{H_t\}$ be the energy process
associated with the collapse process (\ref{eq:16.1}), and {\it
define} the process $\{\xi_t\}$ in terms of the energy process
$\{H_t\}$ and the Brownian motion $\{W_t\}$ according to
\begin{eqnarray}
\xi_t = (T-t)\int_0^t \frac{1}{T-s}(\rd W_s+\sigma_s H_s \rd s).
\end{eqnarray}
Then we have the following:

\vspace{0.15cm} \noindent {\bf Proposition 14}. {\it The random
variables $H_T$ and $\beta_t=\xi_t-\sigma t H_T$ are independent
for all $t\in[0,T]$. Furthermore, the process $\{\beta_t\}$ is a
Brownian bridge.} \vspace{0.15cm}

Proof. For the independence of the random variables $H_T$ and
$\beta_t$ it suffices to verify
\begin{eqnarray}
{\mathbb E}[{\rm e}^{x\beta_t+yH_T}] = {\mathbb E}[{\rm
e}^{x\beta_t}]\,{\mathbb E}[{\rm e}^{yH_T}]
\end{eqnarray}
for arbitrary $x,y$. Using the tower property of conditional
expectation we have
\begin{eqnarray}
{\mathbb E}[{\rm e}^{x\beta_t+yH_T}] = {\mathbb E}\left[ {\rm
e}^{x\xi_t}\,{\mathbb E}\left[{\rm e}^{(y- \sigma tx) H_T}\Big|
\xi_t \right]\right] . \label{eq:19}
\end{eqnarray}
Let us consider the inner expectation ${\mathbb E} \left[ {\rm
e}^{(y- \sigma tx) H_T}|\xi_t\right]$. Using expressions
(\ref{eq:16.6}) and (\ref{eq:16.7}) for the conditional
probability distribution of the terminal energy $H_T$ we deduce
\begin{eqnarray}
{\mathbb E} \left[ {\rm e}^{(y- \sigma tx) H_T}\Big| \xi_t \right]
= \Phi_t^{-1} \sum_i \pi_i {\rm e}^{(y- \sigma tx)E_i} \exp\left(
\frac{\sigma\xi_t E_iT- \frac{1}{2} \sigma^2 E_i^2 t
T}{T-t}\right),
\end{eqnarray}
where the process $\{\Phi_t\}$ is defined in (\ref{eq:17.10}).
Recall now that $\{\Phi_t\}$ is the density process for changing
the measure from ${\mathbb Q}$ to ${\mathbb P}$. As a consequence,
we have
\begin{eqnarray}
{\mathbb E}\left[ {\re}^{x\xi_t}\,{\mathbb E}\left[{\re}^{(y-
\sigma tx) H_T}\Big| \xi_t \right]\right]= {\mathbb E}^{\mathbb Q}
\left[ {\re}^{x\xi_t} \sum_i \pi_i\,{\re}^{(y- \sigma tx)E_i}\,
\re^{\frac{\sigma\xi_t E_iT- \frac{1}{2} \sigma^2 E_i^2 t T}{T-t}}
\right] . \label{eq:19.0}
\end{eqnarray}
However, the process $\{\xi_t\}$ appearing in (\ref{eq:19.0}) is a
Brownian bridge under the ${\mathbb Q}$-measure. Therefore, the
expectation in (\ref{eq:19.0}) can be computed by elementary
methods and we deduce, after some rearrangement of terms, that
\begin{eqnarray}
{\mathbb E}\left[{\re}^{x\beta_t+yH_T}\right] = \sum_i \pi_i\,
{\re}^{yE_i}\,{\re}^{\frac{t(T-t)}{2T}x^2}. \label{eq:22.0}
\end{eqnarray}
Here we have used the facts that if $g$ is a zero-mean Gaussian
random variable with variance $\gamma^2$, then ${\mathbb E}[
{\re}^{xg}]={\re}^{\frac{1}{2}\gamma^2x^2}$, and that the variance
of the ${\mathbb Q}$-Brownian bridge $\{\xi_t\}$ is $t(T-t)/T$.
This proves the independence of $\{\beta_t\}$ and $H_T$. The
result (\ref{eq:22.0}) also establishes that under ${\mathbb P}$
the process $\{\beta_t\}$ is Gaussian, and has mean zero and
variance $t(T-t)/T$. To establish $\{\beta_t\}$ is a Brownian
bridge, we must show that for $s\leq t$ the covariance of
$\beta_s$ and $\beta_t$ is given by $s(T-t)/T$. Alternatively, we
can analyse the moment generating function ${\mathbb E}
[\re^{x\beta_s+y\beta_t}]$. We thus proceed as follows. First,
using the tower property of conditional expectation we have
\begin{eqnarray}
{\mathbb E} \left[\re^{x\beta_s+y\beta_t}\right] &=& {\mathbb E}
\left[\re^{x\xi_s+y\xi_t-\sigma(xs+yt)H_T}\right]\nonumber \\ &=&
{\mathbb E} \left[\re^{x\xi_s+y\xi_t} {\mathbb E}[
\re^{-\sigma(xs+yt)H_T}|\xi_t] \right]\nonumber \\ &=& {\mathbb E}
\left[\re^{x\xi_s+y\xi_t} \Phi_t^{-1}\sum_i\pi_i\,
\re^{-\sigma(xs+yt)E_i}\,\re^{\frac{\sigma\xi_t E_iT- \frac{1}{2}
\sigma^2 E_i^2 t T}{T-t}} \right] \nonumber \\ &=& \sum_i \pi_i \,
\re^{-\sigma(xs+yt)E_i-\frac{tT}{2(T-t)}\sigma^2E_i^2}\, {\mathbb
E}^{\mathbb Q}\left[ \re^{x\xi_s+(y+\frac{\sigma T E_i}{T-t})
\xi_t} \right] . \label{eq:235}
\end{eqnarray}
Now we note
\begin{eqnarray}
{\mathbb E}^{\mathbb Q}\left[ \re^{x\xi_s+(y+\frac{\sigma T
E_i}{T-t})\xi_t}\right] &=& \exp\left\{\half {\mathbb E}^{\mathbb
Q}\left[ \left(x\xi_s+(y+\frac{\sigma E_i T}{T-t})\xi_t
\right)^2\right] \right\} \nonumber \\ && \hspace{-5cm} = \exp
\left\{\half \left( x^2s(1-{\textstyle\frac{s}{T}}) +
(y+{\textstyle\frac{ \sigma E_i T}{T-t}} )^2
t(1-{\textstyle\frac{t}{T}})+2x(y+ {\textstyle\frac{ \sigma E_i
T}{T-t}})s (1-{\textstyle \frac{t}{T}}) \right) \right\},
\end{eqnarray}
from which we observe that the dependence on the energy
eigenvalues $\{E_i\}$ in the summand of (\ref{eq:235}) drops out.
As a consequence, we obtain
\begin{eqnarray}
{\mathbb E}\left[{\re}^{x\beta_s+y\beta_t}\right] = \exp
\left\{\half \left( x^2s(1-{\textstyle\frac{s}{T}}) + y^2
t(1-{\textstyle\frac{t}{T}})+2xys(1-{\textstyle \frac{t}{T}})
\right) \right\}.
\end{eqnarray}
It follows that the covariance of $\beta_s$ and $\beta_t$ for
$s\leq t$ is given by
\begin{eqnarray}
\left.\frac{\partial^2}{\partial x\partial y}{\mathbb E}\left[
{\re}^{x\beta_s+y\beta_t}\right]\right|_{x=y=0}=s\left(
1-\frac{t}{T}\right).
\end{eqnarray}
This establishes the assertion that $\{\beta_t\}$ is a ${\mathbb
P}$-Brownian bridge. \hspace*{\fill} $\diamondsuit$

\section{Time-change and Brownian bridge}
\label{sec:19}

The asymptotic collapse model and the finite-time collapse model
that have been investigated in this paper are in fact related by
an elementary time-change. In this section we shall demonstrate
how a finite-time collapse model can be seen to `emerge' from an
asymptotic collapse model, and vice verse, by the use of
time-change techniques applied to Brownian motion.

We begin by noting the following property of Brownian motion.
Suppose $f(s)>0$ is a continuous monotonic function over
$s\in[0,T]$ such that
\begin{eqnarray}
\int_0^t f^2(s)\, \rd s \to \infty
\end{eqnarray}
as $t\to T$, where $0<T\leq\infty$. Let $\tau(t)$ be given by the
solution of the equation
\begin{eqnarray}
\int_0^{\tau(t)} f^2(s)\, \rd s = t, \label{eq:240}
\end{eqnarray}
or equivalently,
\begin{eqnarray}
f^2(\tau(t))=\left(\frac{\rd\tau}{\rd t}\right)^{-1},
\end{eqnarray}
and let $\{B_t\}$ be a standard Brownian motion. Then the process
$\{X_t\}$ defined by
\begin{eqnarray}
X_t = \int_0^{\tau(t)} f(s)\, \rd B_s  \label{eq:241}
\end{eqnarray}
is a standard Brownian motion. To verify that $\{X_t\}$ is a
Brownian motion it suffices to show that the covariance of $X_s$
and $X_t$ for $s\leq t$ is given by $s$. This follows on account
of the fact that since $\tau(t)$ is deterministic, (\ref{eq:241})
shows that $\{X_t\}$ is Gaussian with mean zero. Using Ito's lemma
we then find that the covariance is given by
\begin{eqnarray}
{\mathbb E}\left[ X_s X_t \right]=\int_0^{\tau(s)} f^2(s)\,\rd s
\label{eq:242}
\end{eqnarray}
for $s\leq t$. Therefore, from the defining relation
(\ref{eq:240}) we conclude that the covariance of $X_s$ and $X_t$
is indeed $s$. As an example, consider the function
\begin{eqnarray}
f(s) = \frac{T}{T-s}. \label{eq:243}
\end{eqnarray}
Clearly $f(s)$ is monotonic and satisfies $f(s)\to\infty$ as $s\to
T$. Furthermore, we have
\begin{eqnarray}
\int_0^\tau f^2(s)\, \rd s = \frac{\tau T}{T-\tau}. \label{eq:244}
\end{eqnarray}
Substitution of (\ref{eq:243}) in (\ref{eq:240}) shows that the
relevant time-change in this example is
\begin{eqnarray}
\tau(t) = \frac{tT}{t+T}. \label{eq:246}
\end{eqnarray}
As a consequence, we find that the process $\{{\tilde B}_t\}$
defined by
\begin{eqnarray}
{\tilde B}_t = \int_0^{\frac{tT}{t+T}} \frac{T}{T-s}\,\rd B_s
\label{eq:time}
\end{eqnarray}
is a standard Brownian motion. From (\ref{eq:246}) we find that
$\tau(0)=0$ and that $\tau(t)\to T$ as $t\to\infty$. Therefore,
the time-change (\ref{eq:246}) has the effect of `slowing down'
the process.

We can consider, conversely, a time-change that has the effect of
`speeding up' the process. For this we require the following
variant of the previous result. Suppose $f(s)>0$ is a continuous
monotonic function over $s\in[0,\infty]$ such that
\begin{eqnarray}
\int_0^t f^2(s)\, \rd s \to T \label{eq:248}
\end{eqnarray}
as $t\to \infty$, where $0<T<\infty$. Let $\tau(t)$ be given by
the solution of the equation (\ref{eq:240}), and let $\{B_t\}$ be
a standard Brownian motion. Then the process $\{Y_t\}$ defined by
\begin{eqnarray}
Y_t = \int_0^{\tau(t)} f(s)\, \rd B_s  \label{eq:249}
\end{eqnarray}
is a standard Brownian motion. This result can be verified by
studying the covariance of $Y_s$ and $Y_t$. As an example we
consider the function
\begin{eqnarray}
f(s) = \frac{T}{s+T},
\end{eqnarray}
which clearly satisfies the condition (\ref{eq:248}). It follows
that the relevant time-change in this example is
\begin{eqnarray}
\tau(t) = \frac{tT}{T-t}, \label{eq:251}
\end{eqnarray}
or equivalently,
\begin{eqnarray}
t = \frac{\tau T}{\tau+T}, \label{eq:253}
\end{eqnarray}
and that the process $\{Y_t\}$ defined by
\begin{eqnarray}
Y_t = \int_0^{\frac{tT}{T-t}} \frac{T}{s+T}\,\rd B_s
\end{eqnarray}
is a standard Brownian motion. We find from (\ref{eq:251}) that
$\tau(0)=0$ and that $\tau(t)\to\infty$ as $t\to T$. Therefore, in
this example the time-change has the effect of speeding up the
clock variable $\tau(t)$.

With these results at hand we now proceed to establish the
relationship between the finite-time and the asymptotic collapse
models studied in this paper. To begin we recall the definition
\begin{eqnarray}
\xi_t=\sigma t H+\beta_t \label{eq:x16}
\end{eqnarray}
for the information process $\{\xi_t\}$ in the finite-time
collapse model, and the fact that the Brownian bridge process
$\{\beta_t\}$ admits the integral representation
\begin{eqnarray}
\beta_t = (T-t) \int_0^t \frac{1}{T-s}\,\rd B_s. \label{eq:x15}
\end{eqnarray}
Next, we consider the time-change given by (\ref{eq:251}) and
define $\{\eta_\tau\}$ by
\begin{eqnarray}
\eta_\tau = \frac{1}{T-t}\,\xi_t . \label{eq:x18}
\end{eqnarray}
Then substituting (\ref{eq:x16}) here and using the integral
representation (\ref{eq:x15}) we obtain
\begin{eqnarray}
\eta_\tau = \frac{\sigma H t}{T-t} + \frac{\beta_t}{T-t} = \sigma
H \tau + \int_0^t \frac{1}{T-s}\,\rd B_s, \label{eq:x19}
\end{eqnarray}
on account of the integral representation (\ref{eq:x15}). However,
recalling the relations (\ref{eq:253}) and (\ref{eq:time}), we
deduce that $\{\eta_\tau\}$ can be expressed in the form:
\begin{eqnarray}
\eta_\tau = \sigma H \tau + {\tilde B}_\tau. \label{eq:x20}
\end{eqnarray}
This defines a `standard' filtering problem associated with the
asymptotic collapse model, if we regard $\tau$ as the time
variable. In particular, the best estimate for $H$, given the
observation $\{{\mathcal F}_\tau^\eta\}$, is determined by the
conditional expectation
\begin{eqnarray}
{\tilde H}_\tau={\mathbb E}[H|\eta_\tau].  \label{eq:x21}
\end{eqnarray}
Clearly we have the relation ${\tilde H}_\tau=H_{t(\tau)}$.
Furthermore, we have
\begin{eqnarray}
\rd {\tilde H}_\tau = \sigma V_\tau \rd {\tilde W}_\tau,
\label{eq:x22}
\end{eqnarray}
where
\begin{eqnarray}
{\tilde W}_\tau = \eta_\tau - \sigma \int_0^\tau {\tilde H}_s \rd
s \label{eq:x23}
\end{eqnarray}
is a standard Brownian motion. From (\ref{eq:x18}) and
(\ref{eq:x23}) we deduce that
\begin{eqnarray}
\rd \left(\frac{1}{T-t}\,\xi_t\right) = \sigma {\tilde H}_\tau
\rd\tau + \rd {\tilde W}_\tau = \sigma H_t \frac{1}{(T-t)^2}\,\rd
t + \frac{1}{T-t}\,\rd W_t,  \label{eq:x24}
\end{eqnarray}
where
\begin{eqnarray}
{\tilde W}_{\frac{tT}{T-t}} = \int_0^t \frac{1}{T-s}\,\rd W_s.
\label{eq:x25}
\end{eqnarray}
Expanding the left side of (\ref{eq:x24}) we deduce
\begin{eqnarray}
\rd\xi_t + \frac{1}{T-t}\Big( \xi_t-\sigma H_t\Big)\rd t = \rd
W_t, \label{eq:x26}
\end{eqnarray}
which, if integrated, reduces to the relation (\ref{eq:17.2}). In
this manner we find that by taking the asymptotic collapse model
(\ref{eq:x20}) and applying the time-change according to
(\ref{eq:251}), which has the effect of `speeding up' the process,
we recover the finite-time collapse model (\ref{eq:x16}).

Conversely, given the model
\begin{eqnarray}
\xi_t = \sigma t H + {\tilde B}_t
\end{eqnarray}
that solves the asymptotic collapse model, we may consider the
time-change given by (\ref{eq:246}) and define
\begin{eqnarray}
\eta_\tau = \frac{T}{t+T}\, \xi_t = \sigma \tau H +
\frac{T}{t+T}\, {\tilde B}_t .
\end{eqnarray}
However, because of the relations $t=\tau T/(T-\tau)$ and
(\ref{eq:time}) we deduce that
\begin{eqnarray}
\frac{T}{t+T}\, {\tilde B}_t = (T-\tau) \int_0^\tau
\frac{1}{T-s}\, \rd B_s,
\end{eqnarray}
which is the integral representation for the Brownian bridge
process $\{\beta_\tau\}$ with respect to the time variable $\tau$.
As a consequence we obtain
\begin{eqnarray}
\eta_\tau = \sigma \tau H + \beta_\tau ,
\end{eqnarray}
and we thus recover the model that solves the finite-time collapse
process. We thus obtain the following conclusion:

\vspace{0.15cm} \noindent {\bf Proposition 15}. {\it The
finite-time collapse model {\rm (\ref{eq:16.1})} can be obtained
from the asymptotic collapse model {\rm (\ref{eq:1.1})} by means
of the time-change defined in {\rm (\ref{eq:251})}. The reverse
transformation is obtained by the time-change defined in {\rm
(\ref{eq:246})}.} \vspace{0.15cm}

\section{Discussion}
\label{sec:20}

The result of Proposition~9 demonstrates that given the dynamical
equation (\ref{eq:1.1}) for the quantum state and the associated
energy process (\ref{eq:1.2}) we can deduce the existence of the
asymptotic random variable $H_\infty$ and an independent noise
process $\{B_t\}$. The variable $H_\infty$ carries the
interpretation of a \emph{hidden variable} in the stochastic
quantum theory. More precisely, because $H_\infty$ is $\{{\mathcal
F}_\infty^W\}$-measurable, its value can only be determined with
certainty after the collapse has taken place. The `quantum noise'
process $\{B_t\}$ represents the `disinformation' that hides
$H_\infty$ before the completion of the collapse process.

Whether the energy-based reduction models considered here suffice
to describe measurements and relaxation phenomena in general in
nonrelativistic quantum mechanics remains an open issue. There are
attempts, for example, to formulate a spontaneous collapse of the
wave packet in a localised region in space (see, e.g.,
\cite{bassi}, for a recent work in this area). However,
localisation of a particle in a small region in space typically
requires large energy. Indeed, in a generic measurement-theoretic
context one requires an infinite amount of energy to confine a
particle in a finite region~\cite{brody5}, and hence it may be
unphysical to speak of a true `position measurement'. Quantities
such as the position or momentum of a particle thus represent what
an experimentalist can \emph{estimate} from appropriate energy
measurements.

This observation is consistent with the point of view put forward
by Wiener~\cite{wiener} that ``under the quantum mechanics, it is
impossible to obtain any information giving the position or
momentum of a particle, much less the two together, without a
positive effect on the energy of the particle examined~$\ldots$
Thus all coupling is strictly a coupling involving
energy~$\ldots$'' The basic idea is that, in order to measure a
physical quantity of a system the measurement apparatus must
interact with the system, and this is achieved in the form of
interchange of particles (typically photons or phonons). When
these particles interact with the measurement apparatus they
create some form of excitation which then allows the device to
estimate quantities of interest. In this regard we can take the
point of view that the energy-based models are of fundamental
importance in describing random phenomenon involving quantum
systems.

\vspace{0.5cm}
\begin{footnotesize}
\noindent DCB acknowledges support from The Royal Society, and LPH
acknowledges the Institute for Advanced Study, Princeton, for
hospitality while part of this work was carried out. We thank
S.~L.~Adler, I.~C.~Constantinou, J.~Dear, and P.~Pearle for useful
comments and discussions.
\end{footnotesize}

\end{document}